\documentclass[manuscript]{aastex}

\slugcomment{Draft version, 2007, May 23}

\shorttitle{JCMT Gould Belt Legacy Survey}
\shortauthors{Ward-Thompson et al.}

\begin{document}

\title{The James Clerk Maxwell Telescope Legacy Survey of
Nearby Star-forming Regions in the Gould Belt}

\author{D. Ward-Thompson\altaffilmark{1}, 
J. Di Francesco\altaffilmark{2},
J. Hatchell\altaffilmark{3}, 
M. R. Hogerheijde\altaffilmark{4},
P. Bastien\altaffilmark{5}, 
S. Basu\altaffilmark{6}, 
I. Bonnell\altaffilmark{7}, 
J. Bowey\altaffilmark{8}, 
C. Brunt\altaffilmark{3}, 
J. Buckle\altaffilmark{9},
H. Butner\altaffilmark{10}, 
B. Cavanagh\altaffilmark{11}, 
A. Chrysostomou\altaffilmark{11,12},
E. Curtis\altaffilmark{9}, 
C. J. Davis\altaffilmark{11}, 
W. R. F. Dent\altaffilmark{13},
E. van Dishoeck\altaffilmark{4},
M. G. Edmunds\altaffilmark{1},
M. Fich\altaffilmark{14},
J. Fiege\altaffilmark{15},
L. Fissel\altaffilmark{16},
P. Friberg\altaffilmark{11}, 
R. Friesen\altaffilmark{2,17}, 
W. Frieswijk\altaffilmark{18},
G. A. Fuller\altaffilmark{19},
A. Gosling\altaffilmark{20},
S. Graves\altaffilmark{9}, 
J. S. Greaves\altaffilmark{7},
F. Helmich\altaffilmark{18},
R. E. Hills\altaffilmark{9}, 
W. S. Holland\altaffilmark{13}, 
M. Houde\altaffilmark{6},
R. Jayawardhana\altaffilmark{16},
D. Johnstone\altaffilmark{2,17},
G. Joncas\altaffilmark{21},
H. Kirk\altaffilmark{2,17}, 
J. M. Kirk\altaffilmark{1},
L. B. G. Knee\altaffilmark{2}, 
B. Matthews\altaffilmark{2}, 
H. Matthews\altaffilmark{22}, 
C. Matzner\altaffilmark{16}, 
G. H. Moriarty-Schieven\altaffilmark{2,11} 
D. Naylor\altaffilmark{23},
D. Nutter\altaffilmark{1},
R. Padman\altaffilmark{9}, 
R. Plume\altaffilmark{24},
J. M. C. Rawlings\altaffilmark{8}, 
R. O. Redman\altaffilmark{2}, 
M. Reid\altaffilmark{25},
J. S. Richer\altaffilmark{9}, 
R. Shipman\altaffilmark{4},
R. J. Simpson\altaffilmark{1},
M. Spaans\altaffilmark{4},
D. Stamatellos\altaffilmark{1},
Y. Tsanis\altaffilmark{8},
S. Viti\altaffilmark{8},
B. Weferling\altaffilmark{11},
G. J. White\altaffilmark{26,27}, 
A. P. Whitworth\altaffilmark{1},
J. Wouterloot\altaffilmark{11},
J. Yates\altaffilmark{8},
M. Zhu\altaffilmark{2,11}}

\altaffiltext{1}{School of Physics \& Astronomy, Cardiff University,
5 The Parade, Cardiff, UK}
\altaffiltext{2}{Herzberg Institute of Astrophysics,National Research 
Council of Canada, 5071 West Saanich Road, Victoria, BC, Canada}
\altaffiltext{3}{School of Physics, University of Exeter, Stocker Road,
Exeter, UK}
\altaffiltext{4}{Leiden Observatory, Leiden University, 
PO Box 9513, 2300 RA, Leiden, The Netherlands}
\altaffiltext{5}{D\'epartement de physique et Observatoire due 
Mont-M\'egantic, Universit\'e de Montr\'eal, C.P. 6128, 
Succ. Centre-ville, Montr\'eal, QC, Canada}
\altaffiltext{6}{Physics and Astronomy Department, 
University of Western Ontario, 1151 Richmond Street, London,
ON, Canada}
\altaffiltext{7}{Scottish Universities Physics Alliance, Physics \& 
Astronomy, University of St Andrews, North Haugh, St Andrews, Fife, UK}
\altaffiltext{8}{Dept of Physics \& Astronomy, University College London, 
Gower Street, London, UK}
\altaffiltext{9}{Cavendish Laboratory, Cambridge University, 
J J Thomson Avenue, Cambridge, UK}
\altaffiltext{10}{Department of Physics and Astronomy, James Madison 
University, 901 Carrier Drive, Harrisonburg, VA 22807, USA}
\altaffiltext{11}{Joint Astronomy Center, 660 N. A'Ohoku Drive,
University Park, Hilo, Hawaii}
\altaffiltext{12}{School of Physics, Astronomy and Mathematics,
University of Hertfordshire, College Lane, Hatfield, UK}
\altaffiltext{13}{UK Astronomy Technology Centre, Royal Observatory, 
Blackford Hill, Edinburgh, UK}
\altaffiltext{14}{Dept of Physics \& Astronomy, University of Waterloo,
Waterloo, ON, Canada}
\altaffiltext{15}{Dept of Physics and Astronomy, 
University of Manitoba, Winnipeg, MB, Canada}
\altaffiltext{16}{Department of Astronomy and Astrophysics, 
University of Toronto, 50 St. George St., Toronto, ON, Canada}
\altaffiltext{17}{Dept of Physics \& Astronomy, University of 
Victoria,  3800 Finnerty Rd., Victoria, BC, Canada}
\altaffiltext{18}{SRON, Netherlands Institute for Space Research,
Landleven 12, 9747 AD Groningen, The Netherlands}
\altaffiltext{19}{School of Physics \& Astronomy, University of Manchester, 
Sackville Street, Manchester, UK}
\altaffiltext{20}{Astrophysics Group, Department of Physics, 
Oxford University, Denys Wilkinson Building, Keble Road, Oxford, UK}
\altaffiltext{21}{Dept de Physique et Observatoire du Mont Megantic, 
Universite Laval, QC, Canada}
\altaffiltext{22}{National Research Council of Canada, Dominion Radio 
Astrophysical Observatory, 717 White Lake Rd., Penticton, BC, Canada}
\altaffiltext{23}{Department of Physics, University of Lethbridge, 
4401 University Dr., Lethbridge, AB, Canada}
\altaffiltext{24}{Dept of Physics \& Astronomy, University of Calgary, 
2500 University Drive, Calgary, AB, Canada}
\altaffiltext{25}{Department of Physics \& Astronomy, McMaster University,
1280 Main St. W., Hamilton, ON, Canada}
\altaffiltext{26}{Dept of Physics and Astronomy,
Open University, Walton Hall, Milton Keynes, UK}
\altaffiltext{27}{Science and Technology Facilities Council,
Rutherford Appleton Laboratory, Chilton, Didcot, UK}


\begin{abstract}

This paper describes a James Clerk Maxwell Telescope (JCMT) legacy survey 
that has been awarded roughly 500 hrs of observing time to be carried 
out from 2007 to 2009. In this survey we will map with SCUBA-2 
(Submillimetre Common User Bolometer Array 2) almost all of
the well-known low-mass and intermediate-mass
star-forming regions within 0.5\,kpc that are accessible from the JCMT.
Most of these locations are associated with the Gould Belt.
From these observations we will  produce a flux-limited snapshot of star 
formation near the Sun,  providing a legacy of images, as well as 
point-source and
extended-source catalogues, over almost 700 square degrees of sky. 
The resulting images will yield the
first catalogue of prestellar and protostellar
sources selected by submillimetre continuum
emission, and should increase the number of known sources 
by more than an order of magnitude.
We will also obtain CO maps with the array receiver
HARP (Heterodyne Array Receiver Programme), 
in three CO isotopologues, of a large typical sample of
prestellar and protostellar sources. 
We will then map the brightest hundred sources with the SCUBA-2
polarimeter (POL-2), producing the first statistically significant set of
polarization maps in the submillimetre. 
The images and
source catalogues will be a powerful reference set for
astronomers, providing
a detailed legacy archive for future telescopes, including ALMA,  
Herschel and JWST.  

\end{abstract}


\section{Introduction}

\subsection{Nearby star formation}

Understanding star formation is a crucial goal of astronomy. Star 
formation plays a pivotal role in most aspects of astronomy from 
the formation and evolution of galaxies to the origins of extra-solar 
planets and the potential for life elsewhere in our Galaxy. Our knowledge 
of the star-formation process has increased dramatically due to the advent 
of sensitive far-infrared and submillimetre
detectors but has suffered from the piece-meal 
fashion in which such observations have been undertaken to date. We
describe here a project that aims to produce a large and 
unbiased sample of star-forming molecular material in the solar vicinity
at relatively high resolution
(8--14 arcsec).

To understand star formation, we need to probe the physical 
conditions of molecular clouds before and during the star formation 
process. Although near-IR images can tell us a great deal about the results 
of star formation, the objects visible are too old to assess the crucial 
conditions in which star formation originates. We need to illuminate 
the earliest conditions in order to understand the formation process
(e.g. Di Francesco et al., 2007; Ward-Thompson et al., 2007).

Submillimetre continuum imaging
selects the very earliest stages of star formation because it traces
the high column densities of dust, even when that dust is
at low temperatures, within
star-forming cores and allows important physical parameters such as
density to be traced in detail.  
From its earliest days the 
James Clerk Maxwell Telescope (JCMT)
has been mapping submillimetre continuum
emission from star-forming regions (e.g. Ward-Thompson et al., 1989; 1995).
The JCMT identified first the youngest
known protostars (Class 0 objects; Andr\'e et al. 1993) and molecular
cloud cores on the verge of collapse to form protostars (prestellar
cores; Ward-Thompson et al. 1994).  

The JCMT also helped determine that
the prestellar core mass function mimics the IMF, indicating that it
may be determined at the very beginning of the star formation process
-- this was
the first observational breakthrough in understanding its
origin in nearly 50 years (Johnstone et al. 2000; Motte et al. 2001;
Nutter \& Ward-Thompson 2007).
Many large area continuum mapping
surveys have also been carried out with 
the Submillimetre Common-User Bolometer Array
(SCUBA) on JCMT
of star-forming regions
(e.g. Johnstone \& Bally 1999; Pierce-Price et al., 2000; 
Johnstone et al., 2000; 2001; 2006;
Hatchell et al., 2005; 2007; Nutter et al., 2005; 2006;
Nutter \& Ward-Thompson 2007; J. Kirk et al., 2005;
Moriarty-Schieven et al., 2006).
Many of these regions lie in a ring around the sky, coincident with the
Gould Belt. 

\subsection{The Gould Belt}

The Gould Belt is a ring of nearby O-type stars inclined at about 20$^\circ$
to the Galactic Plane. It was first discovered in the southern hemisphere
in fact by
John Herschel (1847), who noted that many of the brightest stars in
the southern sky lie in a band that is inclined to the plane of the Galaxy.
Subsequently, Gould (1879) traced the northern part of the band,
thereby completing the ring.

The Gould Belt is
centred on a point $\sim 200\,$pc from the Sun and 
is about 350 pc in radius
(e.g. Clube 1967; Stothers \& Frogel 1974; Comeron et al. 1992;
de Zeeuw et al., 1999; P\"oppel 2001).
Figure~1 shows a schematic of the Gould Belt, and Figure~2
shows its projection onto the plane of the sky,
showing the inclination to the Galactic Plane.

The formation mechanism of the Belt remains 
something of a mystery. One hypothesis is that it
may be the result of a high velocity cloud impacting 
the Galactic Plane
(Comeron \& Torra, 1992; 1994; Guillout et al., 1998).
An alternative possibility is a
local massive supernova remnant or stellar wind interacting
with a large molecular cloud (Blaauw 1991).

Whatever the cause, the Gould Belt is a highly active ring
of nearby star formation, and
most of the local star-forming
molecular clouds are associated with it, including Taurus,
Auriga, Orion, Lupus, Ophiuchus, Scorpius, Serpens and Perseus.
Study of the star formation within the Gould Belt sheds light on the
process of star formation within these respective clouds.
Moreover, it may also help to shed light on the whole of
the Gould Belt itself. For example, accurate dating
of the bursts of star formation around the Belt may be able to test the
various formation mechanisms of the Gould Belt.

\subsection{A new era for JCMT}

The JCMT is currently undergoing a complete overhaul of its instrumentation,
including a new bolometer array camera, 
the Submillimetre Common-User Bolometer Array 2
(SCUBA-2; Holland et al., 2006), 
with an imaging
polarimeter, the SCUBA-2 Polarimeter
(POL-2; Bastien et al., 2005), 
and a heterodyne array receiver, 
the Heterodyne Array Receiver Programme for B-band
(HARP; Smith et al., 2003).
The combination of SCUBA-2, 
HARP, and POL-2
are a powerful tool set with 
which to study star formation.

SCUBA-2 is an innovative 10,000 pixel submillimetre camera due to be 
delivered shortly to the JCMT. The camera is 
expected to revolutionize submillimetre astronomy in terms of its ability 
to carry out wide-field surveys to unprecedented depths.
SCUBA-2 uses Transition Edge Super-conducting
(TES) bolometer arrays, which come complete with in-focal-plane 
Superconducting Quantum Interference Device
(SQUID) amplifiers and multiplexed readouts, and are cooled to 
100mK by a liquid cryogen-free dilution refrigerator.
SCUBA-2 will observe simultaneously at 850 and 450 microns, with
angular resolutions of 14 and 8 arcsec respectively
(Holland et al., 2006).

The polarimeter POL-2 will have an achromatic continuously 
rotating half-wave plate in order to modulate the signal at a rate 
faster than atmospheric transparency fluctuations. Such a modulation 
should improve significantly the reliability and accuracy of submillimetre 
polarimetric measurements. The signal will be analyzed by a wire-grid 
polarizer. For calibration, a removable polarizer will also be available. 
The components, in the order that the radiation will encounter them, are 
the calibration polarizer, the rotating wave plate, and the polarizer. 
The components will be mounted in a box fixed permanently in front of 
the entrance window of the main cryostat of SCUBA-2. All components 
will be mounted so that they can be taken in and out of the beam 
remotely, making it very easy and fast to start polarimetry at the 
telescope (Bastien et al., 2005).

HARP is a 350GHz, 4$\times$4 element, heterodyne focal plane array, using 
SIS detectors, recently commissioned on the JCMT. 
Working in conjunction with the backend
Auto-Correlation and Spectral-line Imaging System (ACSIS;
Hovey et al., 2000),
HARP provides 3-dimensional imaging capability with high sensitivity at 
325 to 375GHz. This is the first submillimetre spectral imaging system on JCMT,
affording significantly improved productivity in terms of speed of mapping. 
The core specification for the array is that the combination of the 
receiver noise temperature and beam efficiency, weighted optimally 
across the array is $<$330K single side-band (SSB) for the central 
20GHz of the tuning range (Smith et al., 2003).
The 16 pixels have receiver temperatures of 94--165~K.
The angular resolution of HARP is 14 arcsec, matching the 850-$\mu$m
resolution of SCUBA-2.

ACSIS has 16 inputs (actually 32, paired up), with a maximum bandwidth 
per channel of approximately 2GHz in a 2$\times$1 GHz configuration.
It has a minimum sample time of 50ms and a maximum output map size of 
16 Gbytes. It has a number of spectral bandwidths and resolutions that
can be selected by the user: 250MHz bandwidth with 30kHz resolution;
500MHz bandwidth with 61kHz resolution 
(multi-subsystem mode); 1GHz bandwidth
with 500kHz resolution; and 2GHz bandwidth with 1000kHz resolution (merged).
In practise, the usable bandwidth will be about 10\% less than this, 
because of the filter roll-off.

ACSIS and HARP together have a number of observing modes,
including raster mapping with position switching for mapping large areas, 
chopped jiggle mapping for fully sampled mapping of areas comparable to 
the HARP focal plane area, and jiggle mapping with fast frequency switching
for fully sampled mapping of compact areas where no nearby off-source 
reference position is available. 

During the planning phase of observations with
the new JCMT instrumentation, numerous ideas
were put forward for science questions that could be addressed 
(e.g. Ward-Thompson 2004). From these plans a number of proposals emerged.
One such proposal, described herein, was  
to map local star-forming regions
with SCUBA-2, HARP and POL-2.
This was one of seven proposals accepted as part of the JCMT Legacy
Programme. 

With the increased mapping speed of SCUBA-2, one can cover
essentially all of the star-forming regions within 0.5\,kpc
in a reasonable amount of time, 
detecting all of the protostars and prestellar cores.  
SCUBA-2 is designed to have
increased sensitivity in each pixel, but also has
two orders of magnitude more pixels than SCUBA, giving it
about a 1000-fold increase in mapping speed.
The increased mapping speed of SCUBA-2 can be illustrated by
comparison with the large-scale mapping survey carried out by SCUBA
of the Galactic Centre (Pierce-Price et al., 2000), which took
$\sim$50 hours of telescope time and covered only 1.4 square degrees. 
By contrast this survey with SCUBA-2 will cover
roughly 700 square degrees to a
greater depth in $\sim$120 hours.

We will use SCUBA-2 to map
the submillimetre continuum emission from as many clouds within 0.5 kpc
as are visible
from the JCMT, including several well known Gould Belt
clouds such as
Orion, Taurus, Perseus, and Ophiuchus.  Several objects
outside of the Gould Belt, including nearby Bok Globules, will also be
mapped (c.f. Launhardt et al., 1997).
We estimate that the source catalogue that we will produce
will contain over five thousand sources. 
Such large
samples of protostars and starless cores are required to provide
robust statistics on objects over a range of evolutionary stages.
With SCUBA-2's predicted improvement in per-pixel
sensitivity over SCUBA, we will measure the 
prestellar clump mass functions down to substellar masses.

HARP increases the heterodyne mapping speed of the JCMT by
over an order of magnitude. 
We will use this increased mapping speed to map a significant fraction of
the SCUBA-2 sources in isotopologues of CO.
The combination
of dust continuum maps from SCUBA-2 plus spectral line data cubes from
the heterodyne array at matched resolution will be extremely powerful.
Since molecular clouds are highly turbulent, and star formation
generates infall, outflow, and rotational motions, velocity
measurements are critical in understanding the mass-assembly process,
feedback, and star-formation efficiency
(e.g. Goodwin et al., 2004a \& b; Vazquez-Semadeni et al., 2005).
In addition, another use of the HARP data will be a determination 
of the amount of line
contamination in the SCUBA data. Johnstone et al. (2003) 
considered this problem
in Orion and argued that it was not significant except for faint sources. 
With the increased sensitivity of
SCUBA2 such contamination may be important.

POL-2 will be able to make
polarization maps of both high and low density material in
molecular clouds.  
We will use it to map one hundred bright sources found in the
SCUBA-2 survey.
At present there is a debate 
over the relative importance of magnetic fields and turbulence 
in regulating the star formation process
(e.g. Mouschovias 1991; Padoan \& Nordlund 2002).
Combined with kinematics from HARP, the POL-2 observations
will allow for an investigation into the balance between gravity,
turbulent support, and magnetic fields over a statistically
meaningful number of star-forming cores. Previously only a few cores have
been mapped (e.g., Holland et al., 1999;
Ward-Thompson et al., 2000; Matthews \& Wilson 2002;
Crutcher et al., 2004; J. Kirk et al., 2006).

The goal of this 
paper is to outline the aims of the
Gould Belt Survey and to describe the 
observations that will be carried out.
In total this programme will take roughly 1000 hours of observing
time on the JCMT, 500 of which have been allocated over the first two years
between 2007/8 and 2009/10. 
Section 2 details the SCUBA-2 aspects of the survey,
section 3 discusses the HARP survey, 
section 4 describes the survey to be carried out with POL-2, 
section 5 outlines some surveys at far-infrared wavelengths being carried out
in parallel with this survey, and
section 6 provides a brief summary of the paper.

\section{SCUBA-2 Survey}

To obtain a complete view of the star formation in the Gould Belt, we
need an inventory of all protostellar objects contained in these clouds.
We will map with SCUBA-2 all of the star-forming 
regions within 0.5 kpc in the Gould Belt accessible by the 
JCMT. The sample, which includes many well-known regions, 
will provide a very significant 
snapshot of star formation near the Sun.  
We will map the thermal dust emission at 850 microns towards the A$_V$
$>$ 1 areas of our target clouds (see Table 1) to a uniform depth,
with a resolution of 14 arcsec. In higher extinction
regions (A$_V$ $>$ 3), we will go deeper, and will utilize better weather
to observe at both 850 and 450 microns -- the latter 
has a superior resolution of 8 arcsec.

Table 1 lists details of
the main clouds to be mapped, and Figures 3 -- 12 
illustrate the approximate mapping areas.
This will provide a legacy
of images, as well as point-source
and extended-source catalogues, of
roughly 700 square degrees of sky.  These maps will be sensitive to every
Class 0 \& I protostar in the Gould Belt
and every L1544-like prestellar core within 200 pc
(J. Kirk et al., 2005).
The maps will yield the first extensive
catalogue of such objects selected by
submillimetre continuum emission and will increase the number of known
sources by more than an order of magnitude.
Comparison with observations taken at other wavelengths, such as 
with Spitzer or Herschel, will allow correct classification
of the sources.

The key science goals of the SCUBA-2 survey are:

\begin{itemize}

\item
to calculate
the duration of each of the protostellar stages;

\item
to elucidate the nature of
the evolution of protostellar collapse;

\item
to discover
the origin of the initial mass
function (IMF) of stars
from intermediate-mass stars to sub-stellar objects;

\item
to discern
the connection between protostars and the molecular cloud structure from
which they formed.  

\end{itemize}

In addition, the SCUBA-2 maps will provide `finding charts' both for
the other aspects of this survey and for future projects. SCUBA-2
will take this subject beyond the source-by-source approach of the
past, into the domain where large-number statistics on the earliest
stages of star formation can finally be carried out, through a wide 
census of starless and prestellar cores and protostars.  

To avoid mapping large areas of blank sky,
molecular clouds have been
pre-selected by visual extinction, $A_V$, from the recent  
extinction atlas of Dobashi et al. (2005).  The continuum mapping will
be divided into two layers, a wide survey of areas with A$_{V}$ = 1-3,
and a deep survey of area with A$_{V} >$ 3.  Figures 3--12 show the extents
of each of the surveys in each region. In addition, some `blank field'
areas will also be mapped as a control sample.

\subsection{Shallow Survey} 

For the shallow survey, we will map areas with $A_V >$
1 at 850\,$\mu$m to a depth of 1 $\sigma$ = 10\ mJy\,beam$^{-1}$.   Within
the Gould Belt, this comprises an area of $\sim$400 square degrees.
In addition, we will map 120 square degrees outside 
of the major star-forming complexes, but positionally 
associated with the Gould Belt. This will
cover nearby small clouds and isolated star formation
regions selected from dark cloud catalogues 
(e.g. Lynds 1962; Cambresy 1999;
Dobashi et al., 2005) and previous catalogues 
(e.g. Clemens \& Barvainis 1988; Jijina et al. 1999;
Lee \& Myers 1999; Visser et al. 2002).

Finally,
we will map $\sim$10 square degrees of blank sky (i.e. $A_{V}$ $<$ 1) 
split into several fields near the Gould Belt to the same depth.
This will act as a control
sample to see if we have missed significant numbers of objects by
using $A_V$ to select our target regions. 
The data from the shallow survey will
be sensitive at 3 $\sigma$ to masses down to the sub-stellar mass limit
of 0.08\ M$_\odot$ per beam for objects at 0.5 kpc that have 
T$_d$ $\geq$ 20\,K, which is
typical of the low extinction parts of molecular clouds.

\subsection{Deep Survey} 

Temperatures vary within molecular clouds and the inner
regions are colder (T$_d$ $\sim$ 10\,K) than the outer regions (T$_d$ 
$\sim$ 20\,K) 
due to increased shielding from the interstellar UV field.  For the 
deep survey
we will map regions with $A_{V}$ $>$ 3 to a depth of 1 $\sigma$ = 3\ 
mJy\,beam$^{-1}$ at 850\,$\mu$m to ensure a complete census of star-forming
cores.  These regions comprise an area of $\sim$64 square degrees.
These data
will also reach the sub-stellar mass limit of 0.08 M$_\odot$ per beam
at 3 $\sigma$ for objects at 0.5\,kpc at T$_d$ = 10\,K.  

Furthermore, these
observations will simultaneously provide maps at 450\,$\mu$m with a mean
1 $\sigma$ rms of 12\ mJy per 450\,$\mu$m beam (i.e. equal to 6\ mJy per
850\,$\mu$m beam after smoothing).  
This is because this aspect of the survey will be carried out in
`Grade 1' weather conditions ($\tau_{\rm 225GHZ} \leq$ 0.05).
The 450\,$\mu$m and 850\,$\mu$m data
together will provide spectral index information, where it is essential
to have comparable resolution, to constrain the dust opacity indices
and thus the masses of the objects.

\subsection{Source Count Predictions} 

The total star formation rate for
clouds within 0.5 kpc is $\sim$6 $\times$ 10$^{-3}\ $M$_\odot$\ yr$^{-1}$
(e.g. McKee \& Williams 1997).  Using the measured IMF (e.g. Kroupa
2001), the total stellar production rate within this distance is
therefore
$\sim$0.02 stars yr$^{-1}$.  The best
current estimates of the timescales for
prestellar cores and Class 0 protostars are $\sim$3 $\times$ 10$^{5}$
yr and $\sim$3 $\times$ 10$^{4}$ yr respectively (e.g. Andr\'e et al., 2000).
Thus, we expect $\sim$6000 prestellar cores and $\sim$600 Class 0 
protostars to be found in our wide survey.  Even allowing for possible
uncertainties in these by factors of $\sim$2, we would still expect 
thousands of objects in total, with hundreds at low mass 
(M $< 0.5$\,M$_\odot$) and tens at high-mass (M $> 8$\,M$_\odot$).
  
These objects will 
fill in the under-populated extremes of currently measured mass functions.  
Note that only tens of
Class 0 protostars and hundreds of prestellar 
cores in total are currently known (Andr\'e et al., 2000).
Finally, the expected numbers of 
objects at the mass function peak (M $\approx 0.5$\,M$_\odot$) will be large
enough ($\sim$20-50 in each cloud) to reveal statistically if
differences in characteristic stellar masses that exist between clouds
are caused by local environmental influences on core formation such as
cloud density and sound speed.

\subsection{Protostellar Lifetimes and Accretion Rates} 

The census of
prestellar cores and protostars from the continuum mapping will allow
us to calculate the relative duration of these stages.  Since half of the
envelope mass is accreted during the Class 0 stage and the rest during
the Class I stage (Andr\'e et al., 1993),
the duration of each stage tells us about the
protostellar accretion rate.   For instance, if they are roughly equal
then the accretion rate is probably constant, as in the Shu collapse
model (Shu 1977).   

Alternatively, if the Class 0 duration is only
one-tenth the Class I duration, as is currently suspected (Andr\'e et
al., 2000), then accretion must start very rapidly and decrease over time
(e.g. Whitworth \& Ward-Thompson 2001), implying a very different collapse
scenario. In addition, if much of the envelope mass is ejected
(e.g. Matzner \& McKee 2000), then the fractions accreted will sum to 
less than 1. Current observations are limited by small-number statistics
(e.g. Visser et al. 2002)
but the wide survey will provide a sufficiently large sample to answer
this question.  

Furthermore, the relative quantities
of prestellar cores at varying degrees of central condensation (given
by continuum radial profiles) will inform models that 
predict the onset of protostellar collapse (e.g.
turbulent dissipation vs. magnetic regulation).

\subsection{Origin of the IMF} 

With the census of nearby
prestellar cores provided by the continuum mapping, we will be able to
plot a very well-populated mass spectrum of these objects over a very
wide range of masses.  This will allow us to confirm or refute the claim
that this mass function (e.g. Motte et al. 2001; Johnstone et al., 2006;
Nutter \& Ward-Thompson 2007) mimics the stellar IMF (Salpeter 1955).
Recent observations appear to show that the prestellar core mass function
follows the same form as the IMF down to very low masses -- see Figure~13
(Nutter \& Ward-Thompson 2007).   

Such a steep mass function implies that low-mass cores dominate (both by
number and by mass). This is in contrast to the cloud mass function where 
most of the mass resides in
the largest structures (e.g. Williams et al., 1994), 
implying that a different 
physical mechanism is responsible for the cores.
Two other obvious differences 
between the cloud and core mass function
also exist. First, while the cloud mass function includes all the mass in 
the cloud (by definition),
the core mass function generally only adds up to a few percent of the mass 
of the parent cloud (Johnstone et al., 2004; H. Kirk et al., 2006).
Second, the mass-radius relation for
clouds implies non-thermal motions for the largest 
structures whereas the cores reveal mostly thermal motions.

If the link between the core mass function and the IMF is
confirmed, this will provide very strong evidence
that the IMF is determined at the prestellar
core stage of star formation -- i.e. the physics behind core formation
is also the cause of the IMF.  The two currently competing theories of
core formation invoke either magnetic fields or turbulence
(e.g. Mouschovias 1991; Ballesteros-Paredes et al., 2003).
Of particular note will be the detection of numerous objects with masses
below the substellar limit (0.08 M$_{\odot}$), allowing for the first
time a comparison between the mass functions of very low mass molecular
cloud cores with the brown dwarf IMF.  
Some detections have already been made of sub-stellar mass cores
(e.g. Greaves et al., 2003), but the numbers are very limited so far.
Similarities between these mass
functions would imply a similar formation process for brown dwarfs and
stars.  A statistically significant deviation, however, would imply
differing origins -- for example,
brown dwarfs may form out of protostellar disks
(for a review, see: Whitworth et al., 2007).

Additional important questions for these very low-mass cores include:
whether they resemble 
the solar mass cores, or
exist only inside a collapsing region;
whether they
also resemble Bonnor-Ebert spheres, and if so, what
provides the confining pressure; or
alternatively, whether they only form within protostellar disks
(e.g. Matzner \& Levin 2005).
Thus the survey has a capacity to constrain brown dwarf formation
directly, as well as through IMF/CMF comparisons.

\subsection{Structure of Cores to Clouds} 

With a distance limit of 0.5\,kpc, the linear 
resolution of the continuum mapping
at 850\,$\mu$m will be 0.03\,pc (7000 AU) or better.  
This scale is well matched for probing the 
structures of the detected prestellar and protostellar envelopes, as 
well as their surrounding environments.  For example, the radial density 
profiles of prestellar cores show
a flat inner region and steep outer region with the turnover at 
$\sim$0.03 pc (Ward-Thompson et al., 1994; J. Kirk et al., 2005).

Some have claimed that the observed radial density profiles
indicate cores are pressure-supported Bonnor-Ebert spheres 
(e.g. Alves et al., 2001), while others have argued
that such configurations result naturally from dynamic evolution of gas 
on large scales and do not indicate equilibrium (Ballesteros-Paredes et al.,
2003). The differences between the model predictions centre around the
velocity profiles of the cores.
The larger sample of cores revealed by the continuum mapping, 
and followed up by CO spectroscopy with HARP (see section 3 below),
will provide enough examples of both velocity
structure and detailed morphology to settle this 
debate. On slightly larger scales, the 
continuum mapping will have the sensitivity to probe the lower-density 
surroundings of these cores, allowing insights into core formation.
For instance, if one assumes that the continuum maps trace the mass, 
then one could use them to assess the
importance of gravitational torques on assembling disks (Jappsen \& 
Klessen 2004).

Previously, these surroundings could be probed only by lower 
density tracers such as CO 
(in regions where it is not depleted and its lines 
are optically thin), but with continuum mapping 
it will be possible to link the structures of cores 
with their environments using the same tracer.  On even larger scales, 
the continuum mapping (especially that of the deep survey) will reveal the 
structure of extended filaments within clouds
(c.f. Johnstone \& Bally 1999; Fiege et al. 2004; Hatchell et al. 2005).

Molecular clouds exhibit filamentary structures, which are often the
locus of star formation along their length. Although turbulence and
magnetic fields have been suggested to explain the filamentary
structures in molecular clouds, few detailed studies have been made
due to limitations in mapping both the velocity fields and the
polarized light which traces the magnetic field. 
Coupled with the dynamical information obtained with 
HARP and the magnetic field geometry obtained with POL-2,
the morphologies and structures of filaments revealed by the continuum
mapping will provide clues both to the origins and evolution of
molecular clouds (c.f. Khalil et al., 2004).

\section{HARP Survey}

In addition to
the SCUBA-2 survey of the local molecular cloud
population,
we will also carry out a CO survey with HARP to trace
the kinematics of cores and clusters,
at the same spatial resolution as the 850-micron SCUBA-2
observations of dust emission (14 arcsec). This
will allow us to address a large number of
fundamental scientific problems in star formation in samples of
statistically significant size for the first time.  

In typical 
star-forming molecular cloud cores, temperatures and densities are in the
ranges 10--50~K and 10$^4$--10$^5$~cm$^{-3}$, which are the conditions
under which the CO and isotopic lines in the 350~GHz range are
excited -- see Table 2.

The key goals of the line observations are: 

\begin{itemize}

\item
to search for and map any high velocity outflows present in the
cores, to differentiate between starless and protostellar
cores;

\item
to derive simple
constraints on the column density and CO depletion in these cores; 

\item
to help understand the support mechanisms and core evolution; 

\item
to characterize the cloud kinematics in a large sample of environments
and investigate the evolution and role of turbulence in star formation.

\end{itemize}

We note that these goals are also very closely coupled to those
of the polarimetric study presented below
and that many of the goals of the continuum survey also
require these data for
interpretation, particularly to classify and determine the
ages of the embedded sources.  

\subsection{Target Regions} 

HARP is a much faster mapping
instrument than previous single-pixel receivers.
However,
it is still not possible to map all of
the regions covered by the SCUBA-2 survey.  Hence we will select
a typical sample of cores and cloud regions from the SCUBA-2 survey,
and observe these with HARP. 

The proposed molecular line observations consist of two sets of
targets. We will make single fully-sampled
footprint images ($\sim$4 square arcmin)
of a set of 1000 cores, and larger
(0.08 degree$^2$) maps of 10 cloud regions containing
filaments and clustered star formation. 
For the well-studied clouds,
many of the target regions are already known from previous SCUBA
surveys.
The cores, selected from the
SCUBA-2 catalogue, will consist of a flux-limited sample plus a
representative selection covering the full range of conditions and
clouds.  

In order to meet the science goals outlined above,
we will map all of the cloud regions and 300 of the cores in
the $^{12}$CO, C$^{18}$O and $^{13}$CO~J = 3--2 lines -- the
latter two lines can be obtained simultaneously within the
ACSIS bandwidth.  A further 700 cores will be mapped in the
$^{12}$CO~3--2 line only. Most of the core maps will be a single
fully-sampled HARP footprint, but for clustered cores we will
combine footprints to map contiguous areas.

As summarised in Table 2, the $^{12}$CO maps will be made with a channel
width of 1.0 km/s.  The target root-mean-square (rms) 1-$\sigma$ noise
level will be 0.3 K in each channel. The  $^{13}$CO and C$^{18}$O 
maps will both
have a channel width of 0.1 km/s, and the target rms 1-$\sigma$ noise
level in each channel will be 0.25 and 0.3 K respectively.  The
slightly poorer predicted rms in C$^{18}$O 
is due to its location near the
edge of an atmospheric band.  The equivalent H$_2$ column density
sensitivities are noted in Table 2.

\subsection{Molecular outflows: classification and mass ejection} 

The first observation
targeting cores will be in the $^{12}$CO\ J = $3-2$ line, to search
for and image any high velocity gas -- see Figure~14.  
The presence of a high velocity
molecular outflow is a standard way to
differentiate between prestellar and protostellar
cores (e.g. Bontemps et al., 1996b;
Wolf-Chase et al., 1998; Visser et al., 2002;
Hatchell et al., 2007).
For compact submillimetre objects with no IR detection, 
outflows are a critical discriminant between
prestellar and protostellar cores.
Where outflows are detected, we will estimate outflow
momentum flux (e.g.,  Bontemps et al., 1996a) in this large homogeneous
sample, allowing us to put on a firm statistical footing the proposed
correlations with bolometric luminosity and envelope mass. 

The envelope mass-luminosity-momentum flux relationships test models of
mass accretion and outflow ejection, and therefore how much of the
envelope mass is ultimately converted into stars
(Matzner \& McKee 2000). Reasonable estimates
of momentum flux can be made without mapping the whole outflow, as
most of the momentum flux is seen close to the star (Fuller \& Ladd
2002).

In the larger cloud regions, mapping the outflows of a large
number of protostellar objects will make it possible to determine
their energy input to the cloud
(e.g.  Norman \& Silk 1980; Li \& Nakamura 2006; Matzner 2007).
The magnitudes of these energies will allow a determination of
the roles of outflows in stimulating turbulent motions or parent 
cloud dispersal (e.g. Silk 1995).

\subsection{CO depletion}

We note that all molecular tracers are affected by chemistry and 
depletion onto dust grains. CO is
no exception, depleting in the densest regions of molecular cores
(e.g. Caselli et al., 1999; Redman et al., 2002).
We will use these observations to
investigate CO depletion. 

Prestellar cores are known to be heavily affected,
with CO depletion factors reaching 10 or more 
(e.g. Crapsi et al., 2005). A
limited study suggests that Class~0 sources are also centrally
depleted by factors of 5, but by the time the Class~I phase is
reached, CO abundances have returned to the canonical
value of $10^{-4}$ (Jorgensen et al., 2002).  

Typical size scales for depletion are $\sim$6000~AU (Caselli et al.,
1999), which corresponds to $12''$ at
500~pc or $40''$ at 150~pc. Outside the density
peaks, CO abundance should be stable (e.g. van Dishoeck 2006; Di
Francesco et al., 2007).

This survey will produce simultaneous dust and CO isotopologue images
of hundreds of cloud cores classified on the basis of infrared
counterparts or molecular outflows.  We can use this sample to
statistically investigate the levels and size scales for CO depletion
as a function of core properties such as mass and evolution.

\subsection{Fundamental properties of cores: mass, temperature and density}

Two independent measurements of the envelope gas mass, to compare with the
dust mass, can be derived from C$^{18}$O. Firstly, the C$^{18}$O
integrated line strength, assuming optically thin emission, will yield
a gas mass.  
Although this will be a lower limit due to depletion it
is still valuable as it relies on a different set of assumptions (CO
abundance, gas excitation) to mass estimates from dust, where the dust
opacity and dust temperature are subject to their own uncertainties.
Secondly, the velocity dispersion and size provide a virial mass
estimate largely unaffected by depletion as linewidths fall towards
cloud cores (e.g. Goodman et al. 1998a \& b). 

In total, the CO observations are vital
to deriving reliable core parameters in order to understand how
clusters form and how the initial mass function is determined.
Combining the HARP data with existing lower-$J$ maps of these
clouds (e.g. Ridge et al., 2006) will
allow us to estimate temperature and density using multi-transition
multi-isotopologue radiative transfer methods.

\subsection{Kinematic tests of star formation models}

Measurement of the detailed
kinematic and density properties of the cores will be carried out using the
C$^{18}$O and $^{13}$CO lines, observable simultaneously
-- see Figure~14. In nearly all cores, we expect
the C$^{18}$O line to be optically thin. Hence, we will be able to
measure the thermal and non-thermal contributions to the line width,
enabling a study of core support mechanisms
and their evolution (e.g. Jessop \& Ward-Thompson 2001).

The combination of a high-density tracer such as the J=3--2
transition, with lower-density tracers (e.g. lower-J CO lines) allows an
investigation of the dynamical motion of cores through their parent
cloud (e.g. Walsh et al., 2004; 2006; Ayliffe et al., 2007).

Where cores are clustered, we will
also compare the velocity dispersion between
cores, a simple quantity to measure which can be compared both with
star formation simulations and with young clusters of
pre-main-sequence stars (e.g. Belloche et al., 2001).

These quantities may potentially discriminate
between competitive accretion/core mergers
(e.g. Bate et al., 2003) and the alternative
where one submillimetre core results in one star system
(e.g. Goodwin et al., 2004a,b).

In addition, the distribution of C$^{18}$O line velocities of cores 
observed with HARP
in individual molecular clouds will probe theories of cloud 
turbulence. Such
cores may act as test particles within the larger-scale 
turbulent motions within
such clouds. Similar smaller-scale studies have recently been 
carried out in, for example,
Perseus (H. Kirk et al., 2007).

\subsection{Clouds and Filaments} 

We will also produce large maps of 10 filamentary or clustered star
formation regions.  Many of these will be well-studied regions (e.g. the
Orion filament and the $\rho$ Ophiuchi cluster), but not all of them
have pre-existing submillimetre maps (e.g. the Pipe Nebula and L1506 in
Taurus).  These regions will be mapped in 
$^{12}$CO, C$^{18}$O and $^{13}$CO lines to
the spectral resolutions summarized in Table 2.  These maps will
cover regions of order 300 square arcminutes (0.08 deg$^2$) in
size.

The key
goal here will be to compare the observations to gas dynamic and MHD
turbulence simulations (e.g. Balsara et al., 2001; Padoan \& Nordlund
2002; Bate et al. 2003; Vazquez-Semadeni et
al. 2005).  These models make strong predictions on the expected clump
mass spectrum, the spatial and velocity structure, the shapes of the
cores and filaments, and the star formation efficiency. 

Comparisons
can be made by analysing cloud spatial and velocity structure using
techniques such as clumping analyses (Williams et al., 1994), axial
ratios (e.g. Kerton et al., 2003), delta variance (Stutzki et al. 1998;
Bensch et al. 2001),
principal component analysis (e.g. Heyer
\& Schloerb 1997), structure
function (e.g. Brunt et al., 2003), and spectral correlation function
(Rosolowsky 1999).

Mapping the J = 3--2 transitions, which trace
reasonably high densities, at an angular resolution typically less 
than a thermal
Jeans length (for an H$_2$ number density of $10^5\,$cm$^{-3}$), reveals
the kinematics over the critical transition to gravitational dominance
(Ossenkopf 2002), whereas existing low J CO maps (e.g. Dame et al.,
2001; Wilson et al., 2005) trace only the lower density large-scale 
molecular cloud structure and kinematics. 
We note that lower J (usually J=1--0) maps exist for many of our
clouds and can be analysed in conjunction with the HARP data
(e.g. Ridge et al., 2006).

In total, the CO observations are vital in deriving reliable core parameters 
in order to understand how clusters form and how the initial mass
function is determined.  
When combined with the SCUBA-2 data, the HARP
data will provide a powerful way to explore the detailed dynamics and
density structure in star-forming clouds, presenting rigorous tests of
the details of the theoretical models. 

\section{POL-2 Survey}

We will also carry out a polarimetric survey of the brightest
cores and clouds, since
measurements of polarized emission
from dust are the most effective means of probing the magnetic field.
By contrast, absorption polarimetry is
limited to the periphery of dense clouds, and Zeeman splitting
detections are made in only a small fraction of
regions observed (Crutcher 1999).  

Polarized dust
emission is detected from objects on all scales measured (e.g.
Matthews et al. 2001; Crutcher et al. 2004) and from compact
cores, regardless of evolutionary epoch (e.g. Ward-Thompson et al. 2000;
Matthews \& Wilson 2002), as illustrated in 
Figure~15.  
Sensitivity limitations and practical limits on observing times 
have previously
restricted observations to several dozen bright 
(i.e. S$_{850\mu m} \geq 1$\,Jy), compact objects.   

Attempts to measure the magnetic field 
in star-forming regions are driven by the need to understand its
significance in the formation of cloud structure and/or the regulation
of collapse of cloud cores.  These factors are related to star
formation rates and molecular cloud lifetimes, on which there is
substantial debate in the literature (e.g. Myers \& Goodman 1988; Hartmann
et al.\ 2001; Elmegreen 2000).

The key goals of the polarimetric mapping are: 

\begin{itemize}

\item
to obtain maps of polarization position angle and fractional
polarization in a statistically meaningful sample of cores;

\item
to characterize the evidence for, and the relevance of, the field
and turbulence (in conjunction with HARP observations) in cores
and their surrounding environments;

\item
to test the predictions of low-mass star formation
theories (core, outflow, field geometry);

\item
to generate a large sample of cores suitable for follow-up with 
forthcoming instruments such as ALMA.

\end{itemize}

\subsection{Target regions} 

From the SCUBA-2 850\,$\mu$m continuum survey
we expect to detect at least 100 cores of sufficient
brightness ($>$250 mJy beam$^{-1}$ peak 850\,$\mu$m flux) for
observation with POL-2 in single SCUBA-2 footprints. These
polarimetry targets will be located in different star-forming regions
and be of various classes (starless, prestellar, Class 0, Class I).  

To probe the
initial field conditions for core formation, maps of $\sim$0.08 deg$^2$
in extent will be made of 10 cloud regions containing
filaments and clustered star formation in synergy with the HARP
large-scale mapping to measure the gas dynamics and
characterize the turbulence. 

Candidate types of regions would
include a starless fragment of a filamentary cloud, such as a portion
of the Pipe Nebula (see Figure~9). In addition we will map
filamentary clouds for which absorption
polarimetry exists, such as the L1506 filament in Taurus.
We will use these data to
compare directly the magnetic field directions yielded by
emission and absorption polarimetry at the same locations, 
allowing these methods to be directly compared for the first time.
We will also map regions
around clusters exhibiting different rates of star formation.

\subsection{Tests of the Models} 

According to one
paradigm of low-mass star formation, collapse is guided by
magnetic fields, producing flattened cores and disks
(e.g. Mouschovias 1991). Outflows are
then generated orthogonal to the disk, producing outflows aligned with
the nascent field direction.  Greaves et al. (1997) tested this
correlation for five sources, finding the alignment of field and
outflow appeared to be correlated to the angle of the outflow to the
line of sight (see also Matthews et al. 2007).

With $\sim$ 70 sources with outflows (Class 0 and
Class I), we will be able to establish statistically whether outflows
are preferentially oriented with respect to the field direction.
Using the SCUBA-2 data, we will also determine whether the field
direction is related to the core morphology.  
Many cores for which polarization maps
exist are too distant to compare field geometry to core morphology,
since the cores are poorly resolved.  A sizeable fraction of the
objects mapped are also embedded in filaments, making measurements of
core axis ratios particularly challenging.

Models predict different
relations between core morphology and polarization position angle
depending on field geometry (straight or helical) and core morphology
(oblate, prolate or triaxial). Since the outflow, core and
field orientations are all measured in 2D projection, statistical
corrections can be made for an ensemble of cores which are not
applicable to individual objects.

\subsection{Models of Magnetic Field Geometry} 

Polarized
dust emission yields only the 2D field geometry projected onto the
plane of the sky.  Utilizing all three components from the survey
(continuum, line and dust polarimetry), we will generate a set of
analytic models of the three dimensional field geometry of all 100
cores as well as extended coherent structures such as filaments.  
This will be practical utilizing a new generalized
modelling code (Fiege 2005), which has already been applied to the case
of a filamentary cloud (Fiege et al., 2004).  

The
most general question to be addressed is whether there are any
quasi-static magnetic models of cores that can explain the
polarization and continuum maps for a statistically significant number
of sources, as some theoretical models predict (e.g.  Mouschovias
1976; Tomisaka et al. 1988). 

If all quasi-static models fail to provide adequate fits to the data, then
the data could only be explained by MHD turbulence.  Regardless of the
outcome, the polarization dataset will settle this important question
in star formation theory.  It will also provide a polarization
catalogue of legacy value, which can be used as a definitive test for
future models and simulations, and a powerful tool for their
refinement. 

The resulting modelling database will provide the theoretical
counterpart to the observational dataset, which would be augmented as
new and improved models are developed.  In principle, one could search
for regions in parameter space where the allowed solution sets
intersect for multiple cores.  This would provide a way to search for
preferred ranges of parameters, which would help to refine future
models.

\subsection{Magnetic Field Strength}

The fractional polarization from dust
yields no direct estimate of the magnetic field strength, since it is
dependent on several additional unknowns (e.g. degree of grain alignment,
grain shape and composition).  The field strength will be derived from
the commonly-used Chandrasekhar-Fermi (CF) method (Chandrasekhar \& Fermi
1953) utilizing dispersion in polarization vectors (where high
dispersion indicates a highly turbulent field and a weak mean field
component), the line widths estimated from the HARP C$^{18}$O/$^{13}$CO 
data, and the density from the SCUBA-2 fluxes
(c.f. Crutcher et al., 2004; J. Kirk et al., 2006).

Simulations show that this
estimate can be corrected for a statistical ensemble of objects to
yield realistic estimates of the field strength (Ostriker et al.,
2001; Heitsch et al. 2001).  Due to measurements of line widths
and dispersion estimates, we will be able to test effectively whether
the cores with broader line widths show more dispersion, thereby
determining the applicability of the CF method in molecular clouds.

\subsection{Large-Scale Fields and Turbulence} 

The same
extended regions will be mapped with POL-2 and HARP.
The relation between the core field geometry and
that of the larger-scale structure in clustered and filamentary
regions will be observed and modelled. Polarization maps can reveal
abrupt changes in polarization direction (see Figure~15), which
indicate underlying changes in the magnetic field direction.  

The gas dynamics from 
HARP C$^{18}$O/$^{13}$CO observations will be used with the polarization
maps to test predictions of magnetized simulations of turbulence
(e.g., Heitsch et al.\ 2001; Ostriker et al. 2001; Padoan \& Nordland
2002).  Such large-scale maps are the key to unraveling the overall
geometry of the magnetic field.

\section{Other Surveys}

In parallel with this survey, there are a number of far-infrared surveys
also being carried out of the Gould Belt. The two most relevant
to this paper are those being carried out on the Spitzer Space Telescope and
the Herschel Space Telescope.

The Spitzer Space Telescope -- formerly SIRTF, the Space Infrared Telescope 
Facility (Werner et al., 2004)
-- was launched on 25 August 2003. It
consists of an 85-cm telescope and three cryogenically-cooled science 
instruments, covering a wavelength range of 3--180$\mu$m.
The three instruments are
the Infrared Array Camera (IRAC; Fazio et al., 2004), 
the Infra-red Spectrograph (IRS; Houck et al., 2004), and
the Multi-band Imaging Photometer for Spitzer (MIPS; Rieke et al., 2004).
IRAC is a four-channel camera that provides simultaneous 
5.12 $\times$ 5.12 arcmin 
images at 3.6, 4.5, 5.8, and 8 microns. Each of the four detector arrays in 
the camera are 256 $\times$ 256 pixels in size. 
The IRS has four separate modules: a low-resolution, short-wavelength mode 
covering the 5.3-14 micron interval; a high-resolution, short-wavelength 
mode covering 10-19.5 microns; a low-resolution, long-wavelength mode for 
observations at 14-40 microns; and a high-resolution, long-wavelength mode 
for 19-37 microns. MIPS consists of
a 128 $\times$ 128 array for imaging at 24 microns, a
32 $\times$ 32 array for imaging at 70 microns, 
and a 2 $\times$ 20 array for imaging 
at 160 microns.

One legacy programme on Spitzer is currently mapping the
same Gould Belt clouds as are described in this paper using IRAC and MIPS
-- for more information see: $http://www.cfa.harvard.edu/gouldbelt/$.
This is a follow-up to the `Cores to disks' Spitzer Legacy Project (c2d;
Evans et al. 2003; see also: Allen et al. 2007; Guedel et al. 2007)

The Herschel Space Telescope (Pilbratt, 2005a,b) is
a 3.5 metre diameter passively cooled telescope due to be launched in 2008.
The science payload complement comprises: the
Heterodyne Instrument for the Far Infrared (HIFI;
de Graauw et al., 2005), a very high resolution 
heterodyne spectrometer; the
Photo-detector Array Camera and Spectrometer (PACS;
Poglitsch et al., 2005), an imaging photometer 
and medium-resolution grating spectrometer, and the
Spectral and Photometric Imaging Receiver (SPIRE;
Griffin et al., 2002), an imaging photometer 
and an imaging Fourier transform spectrometer.

One Guaranteed Time Programme on Herschel will use
PACS and SPIRE to map the same clouds in the Gould Belt at
wavelengths of 75, 170, 250, 350 and 500$\mu$m.
For a full description of this programme see:
$http://starformation-herschel.iap.fr/$.

\section{Summary}

We have presented a programme of observations that will be carried out by 
the JCMT using its new suite of instruments SCUBA-2, HARP and POL-2.
These instruments will be used to survey the nearby star-forming regions
in the Gould Belt to answer key questions in star formation research.
The results of the survey will include legacy images 
and submillimetre source catalogues of the mapped
regions. The data will be presented in a series of data papers
followed by a number of papers interpreting the data and
addressing the science questions discussed above.
The combination of this survey ,
together with those mentioned in the previous section,
will provide a unique set of complementary
data, and a powerful legacy for many years to come.

\newpage

{\bf FIGURE CAPTIONS}

{\bf Fig.1.}
Diagram illustrating the Gould Belt and the positions of each
of its constituent clouds relative to the Sun. The shaded area represents 
that part of the Belt not visible from the JCMT.

{\bf Fig.2.}
All-sky image of the IRAS 100-$\mu$m data plotted in Galactic coordinates,
showing the projection of the Gould Belt onto the plane of the sky and the
positions of the various clouds.

{\bf Fig.3.}
Extinction map of Orion, showing the approximate area it is planned to map
with SCUBA2. The area surrounded by the dark box
will be mapped in the wide shallow 
survey and the area surrounded by the white box
will be mapped in the narrow, deep survey.

{\bf Fig.4.}
Extinction map of Taurus. 
Details as in Figure~3.

{\bf Fig.5.}
Extinction map of Auriga. 
Details as in Figure~3.

{\bf Fig.6.}
Extinction map of Perseus. 
Details as in Figure~3.

{\bf Fig.7.}
Extinction map of Cepheus. 
Details as in Figure~3.

{\bf Fig.8.}
Extinction map of IC5146. 
Details as in Figure~3.

{\bf Fig.9.}
Extinction map of Serpens. 
Details as in Figure~3.

{\bf Fig.10.}
Extinction map of the Pipe Nebula. 
Details as in Figure~3.

{\bf Fig.11.}
Extinction map of Oph-Sco. 
Details as in Figure~3.

{\bf Fig.12.}
Extinction map of Lupus. 
Details as in Figure~3.
Note that Lupus III, IV \& VI are too far south to map with JCMT.

{\bf Fig.13.}
Extinction map of Corona Australis. 
Details as in Figure~3.

{\bf Fig.14.}
The core mass function (CMF) in Orion, from Nutter \& Ward-Thompson 2006.
A three-part stellar IMF, normalised to the peak of the CMF,
is overlaid as a thin solid line. The dotted line shows a three-part mass
function, with the same slopes as the IMF, superimposed on the CMF.
The agreement between the two appears to indicate that the stellar IMF is 
determined by the CMF.

{\bf Fig.15.}
The Class~0 source L1157 mapped in
C$\,^{18}$O~3--2 (greyscale and thin contours) 
and $^{12}$CO~3--2 (thick contours).

{\bf Fig.16.}
SCUBA polarisation maps.
These images show the importance of
mapping the ambient environment around cores and the advantage of high
resolution.
(a) The SCUBA Serpens map (Davis et al. 2000) shows that the
field structure can be more complex between the cores than their
polarization patterns indicate. 
(b) The same is true in Barnard 1 at centre
(Matthews \& Wilson 2002). 
(c) NGC 2024 shows that SCUBA
observations (dark grey vectors) 
of extended structure in Orion (Matthews, Fiege \&
Moriarty-Schieven 2002) reveal systematic variations missed in earlier
lower resolution maps (light grey vectors; Dotson et al. 2000).  This map
reveals how critical resolution is to the effective mapping of
magnetic field geometry.

\newpage

\begin{table}
\begin{tabular}{lcccccc}
Region 		& \multicolumn{3}{c}{Shallow Survey Details} & 
\multicolumn{3}{c}{Deep Survey Details} \\
Name 		& \multicolumn{2}{c}{Field centre} & 
Area & \multicolumn{2}{c}{Field centre} & Area \\
 & RA(2000) &  Dec (2000) & Degree$^2$ &
   RA(2000) &  Dec (2000) & Degree$^2$ \\
		& 	& 	& 	& \\
Orion 		& 05$^{\rm h}$40$^{\rm m}$00$^{\rm s}$ &  
$-$02$^{\circ}$00$^{\prime}$00$^{\prime\prime}$ & 81.1	
& 05$^{\rm h}$40$^{\rm m}$00$^{\rm s}$ &  
$-$02$^{\circ}$00$^{\prime}$00$^{\prime\prime}$	& 14.3 \\
Taurus	 	& 04$^{\rm h}$32$^{\rm m}$00$^{\rm s}$ &  
$+$26$^{\circ}$10$^{\prime}$00$^{\prime\prime}$	& 45.5	
& 04$^{\rm h}$32$^{\rm m}$00$^{\rm s}$ &  
$+$26$^{\circ}$10$^{\prime}$00$^{\prime\prime}$	& 4.5 \\
Auriga		& 04$^{\rm h}$20$^{\rm m}$00$^{\rm s}$ &  
$+$38$^{\circ}$05$^{\prime}$00$^{\prime\prime}$	& 60.3	
& 04$^{\rm h}$31$^{\rm m}$00$^{\rm s}$ &  
$+$36$^{\circ}$40$^{\prime}$00$^{\prime\prime}$	& 2.5 \\
Perseus 	& 03$^{\rm h}$37$^{\rm m}$00$^{\rm s}$ &  
$+$31$^{\circ}$15$^{\prime}$00$^{\prime\prime}$	& 15.5	
& 03$^{\rm h}$37$^{\rm m}$00$^{\rm s}$ &  
$+$31$^{\circ}$15$^{\prime}$00$^{\prime\prime}$	& 2.3\\
Cepheus 	& 21$^{\rm h}$20$^{\rm m}$00$^{\rm s}$ &  
$+$72$^{\circ}$30$^{\prime}$00$^{\prime\prime}$	& 37.7	
& 21$^{\rm h}$20$^{\rm m}$00$^{\rm s}$ &  
$+$72$^{\circ}$30$^{\prime}$00$^{\prime\prime}$	& 1.0\\
IC5146 		& 02$^{\rm h}$48$^{\rm m}$00$^{\rm s}$ &  
$+$47$^{\circ}$30$^{\prime}$00$^{\prime\prime}$	& 0.8	
& 02$^{\rm h}$48$^{\rm m}$00$^{\rm s}$ &  
$+$47$^{\circ}$30$^{\prime}$00$^{\prime\prime}$	& 0.5\\
Serpens 	& 18$^{\rm h}$23$^{\rm m}$00$^{\rm s}$ &  
$-$03$^{\circ}$10$^{\prime}$00$^{\prime\prime}$	& 29.1	
& 18$^{\rm h}$23$^{\rm m}$00$^{\rm s}$ &  
$-$03$^{\circ}$10$^{\prime}$00$^{\prime\prime}$	& 14.7\\
Pipe 		& 17$^{\rm h}$31$^{\rm m}$00$^{\rm s}$ &  
$-$26$^{\circ}$00$^{\prime}$00$^{\prime\prime}$	& 16.4	
& 17$^{\rm h}$31$^{\rm m}$00$^{\rm s}$ &  
$-$26$^{\circ}$00$^{\prime}$00$^{\prime\prime}$	& 7.2\\
Ophiuchus 	& 16$^{\rm h}$26$^{\rm m}$00$^{\rm s}$ &  
$-$24$^{\circ}$30$^{\prime}$00$^{\prime\prime}$	& 30.4	
& 16$^{\rm h}$26$^{\rm m}$00$^{\rm s}$ &  
$-$24$^{\circ}$30$^{\prime}$00$^{\prime\prime}$	& 6.5\\
Scorpius 	& 16$^{\rm h}$51$^{\rm m}$00$^{\rm s}$ &  
$-$25$^{\circ}$20$^{\prime}$00$^{\prime\prime}$	& 31.6	
& 16$^{\rm h}$51$^{\rm m}$00$^{\rm s}$ &  
$-$25$^{\circ}$20$^{\prime}$00$^{\prime\prime}$	& 1.4\\
Lupus I		& 15$^{\rm h}$40$^{\rm m}$00$^{\rm s}$ &  
$-$34$^{\circ}$30$^{\prime}$00$^{\prime\prime}$ & 23.3	
& 15$^{\rm h}$40$^{\rm m}$00$^{\rm s}$ &  
$-$34$^{\circ}$30$^{\prime}$00$^{\prime\prime}$	& 3.5\\
Lupus II	& 16$^{\rm h}$00$^{\rm m}$00$^{\rm s}$ &  
$-$38$^{\circ}$00$^{\prime}$00$^{\prime\prime}$ & 1.7	
& 16$^{\rm h}$00$^{\rm m}$00$^{\rm s}$ &  
$-$38$^{\circ}$00$^{\prime}$00$^{\prime\prime}$	& 0.2\\
Lupus V		& 16$^{\rm h}$20$^{\rm m}$00$^{\rm s}$ &  
$-$37$^{\circ}$30$^{\prime}$00$^{\prime\prime}$ & 15.7	
& 16$^{\rm h}$20$^{\rm m}$00$^{\rm s}$ &  
$-$37$^{\circ}$30$^{\prime}$00$^{\prime\prime}$	& 2.4\\
CrA 		& 19$^{\rm h}$15$^{\rm m}$00$^{\rm s}$ &  
$-$37$^{\circ}$30$^{\prime}$00$^{\prime\prime}$	& 11.0	
& 19$^{\rm h}$02$^{\rm m}$00$^{\rm s}$ &  
$-$37$^{\circ}$00$^{\prime}$00$^{\prime\prime}$	& 2.8\\
 		& 	& 	& 	& \\
\end{tabular}
\caption{Technical details of approximate areas to be mapped with SCUBA-2.}
\end{table}

\newpage

\begin{table}

\begin{center}

\begin{tabular}{cccccc}
Transition              &$\nu$    &$E_U/ k$ &$\Delta v$ &$T_{\mathrm RMS}$  
&$N_{H_2} (3\sigma)$\\
                        &GHz    &K            &km s$^{-1}$  &K   
&cm$^{-2}$\\
 & & & & & \\
$^{12}\hbox{CO }J=3\hbox{--}2$ &345.7960 &33.2 &1.0  &0.3         
&$4.5\times 10^{18}$ \\
$^{13}\hbox{CO }J=3\hbox{--}2$ &330.5880 &31.7 &0.1  &0.25        
&$3.6\times 10^{19}$ \\
C$^{18}$O~$J=3\hbox{--}2$ &329.3305 &31.6 &0.1  &0.3         
&$3.3\times 10^{20}$ \\
 & & & & & \\
\end{tabular}

\end{center}

\caption{CO isotopologues to be observed with HARP.
Column density calculations assume LTE at 50~K  for $^{12}$CO
and 20~K 
for $^{13}$CO \& C$^{18}$O. Abundances relative to H$_2$ 
are taken to be 10$^{-4}$
for $^{12}$CO and $1.7\times 10^{-7}$ for C$^{18}$O (Frerking et al. 1982).
The ratio of
$^{13}\hbox{CO}/\hbox{C}^{18}\hbox{O}$ is assumed to be 77
(Wilson \& Rood 1994). The
standard conversion of column density to visual extinction is
taken, such that an $A_v$ of 1 corresponds to a column density of 
$0.9\times 10^{21} (N_{H} + 2N_{H{_2}})$~cm$^{-2}$.}
\label{colines}

\end{table}

\newpage

\begin{figure}
\setlength{\unitlength}{1mm}
\begin{picture}(120,120)
\includegraphics{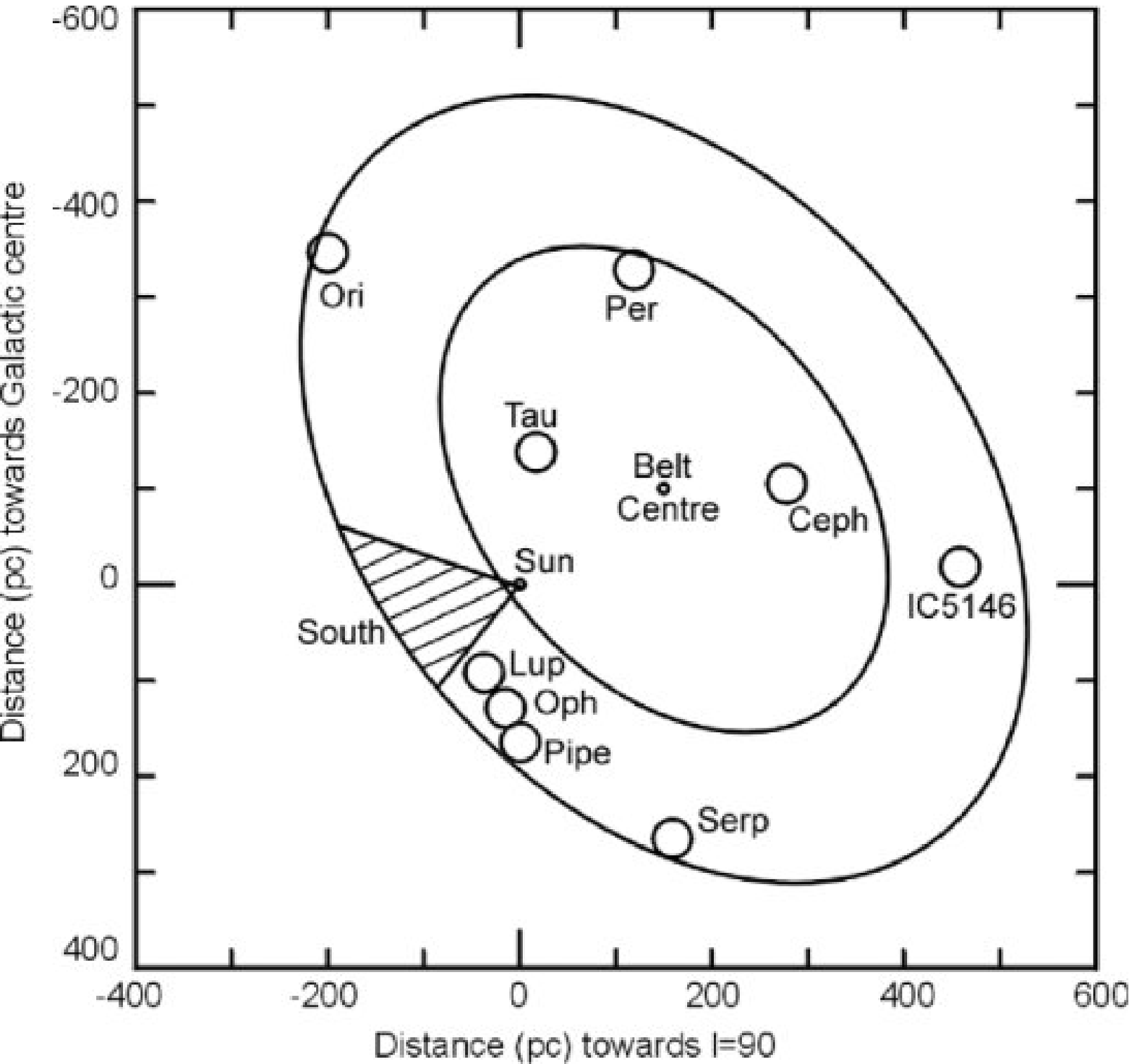}
\vspace*{20cm}
\end{picture}
\caption{}
\label{fig1}
\end{figure}

\newpage

\begin{figure}
\setlength{\unitlength}{1mm}
\begin{picture}(120,120)
\includegraphics{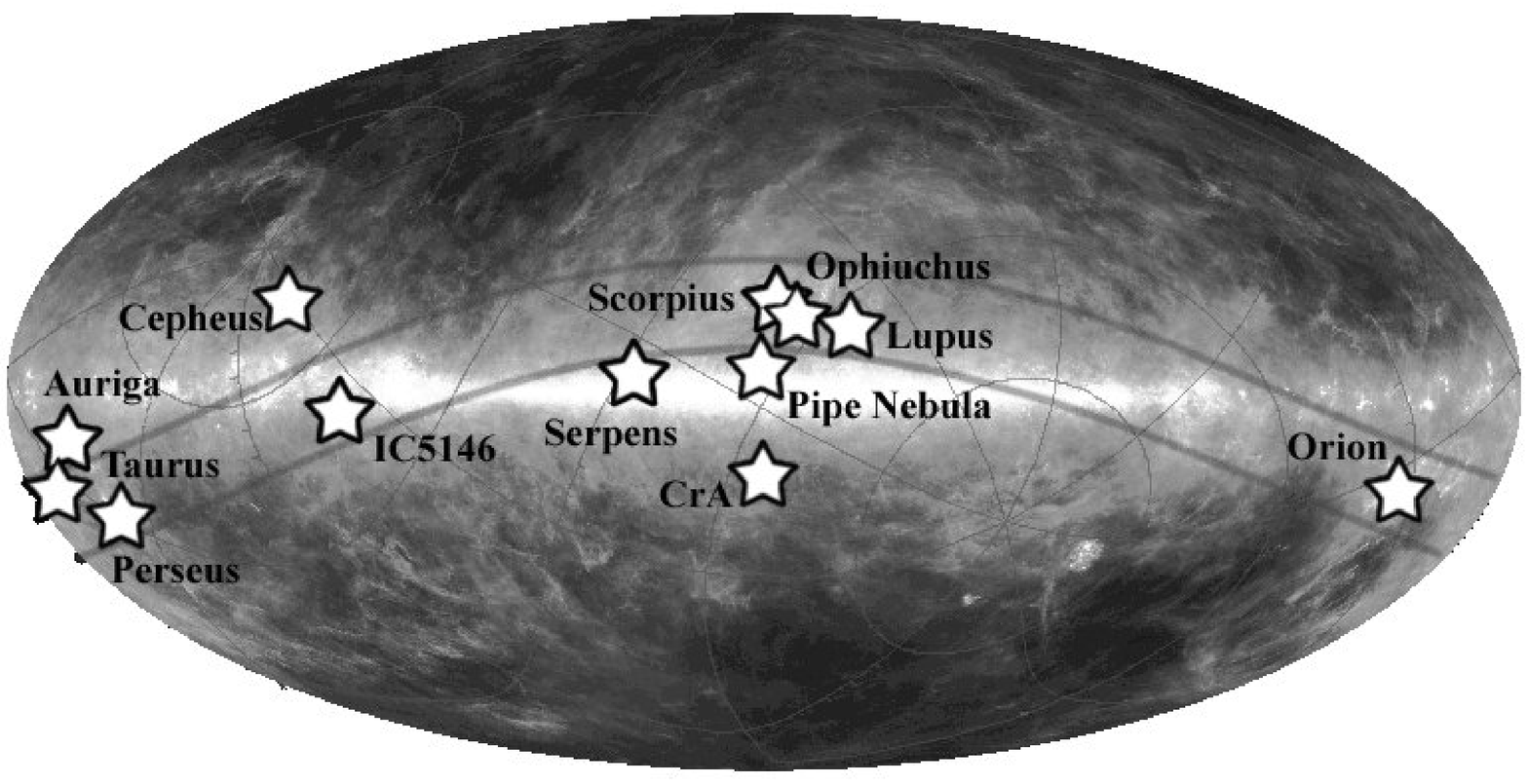}
\vspace*{20cm}
\end{picture}
\caption{}
\label{fig2}
\end{figure}

\newpage

\begin{figure}
\setlength{\unitlength}{1mm}
\begin{picture}(120,120)
\includegraphics{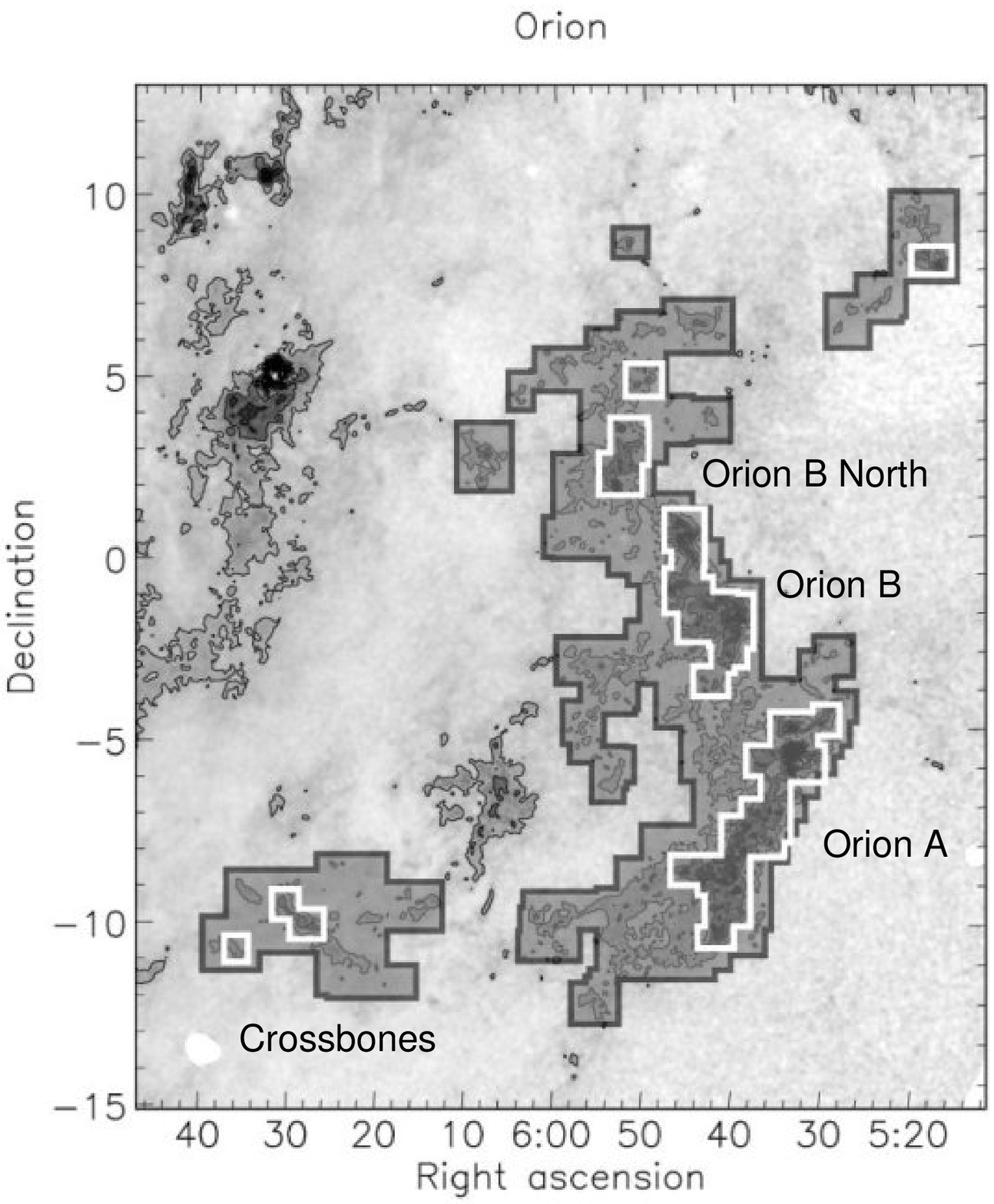}
\vspace*{20cm}
\end{picture}
\caption{}
\label{fig3}
\end{figure}

\newpage

\begin{figure}
\setlength{\unitlength}{1mm}
\begin{picture}(120,120)
\includegraphics{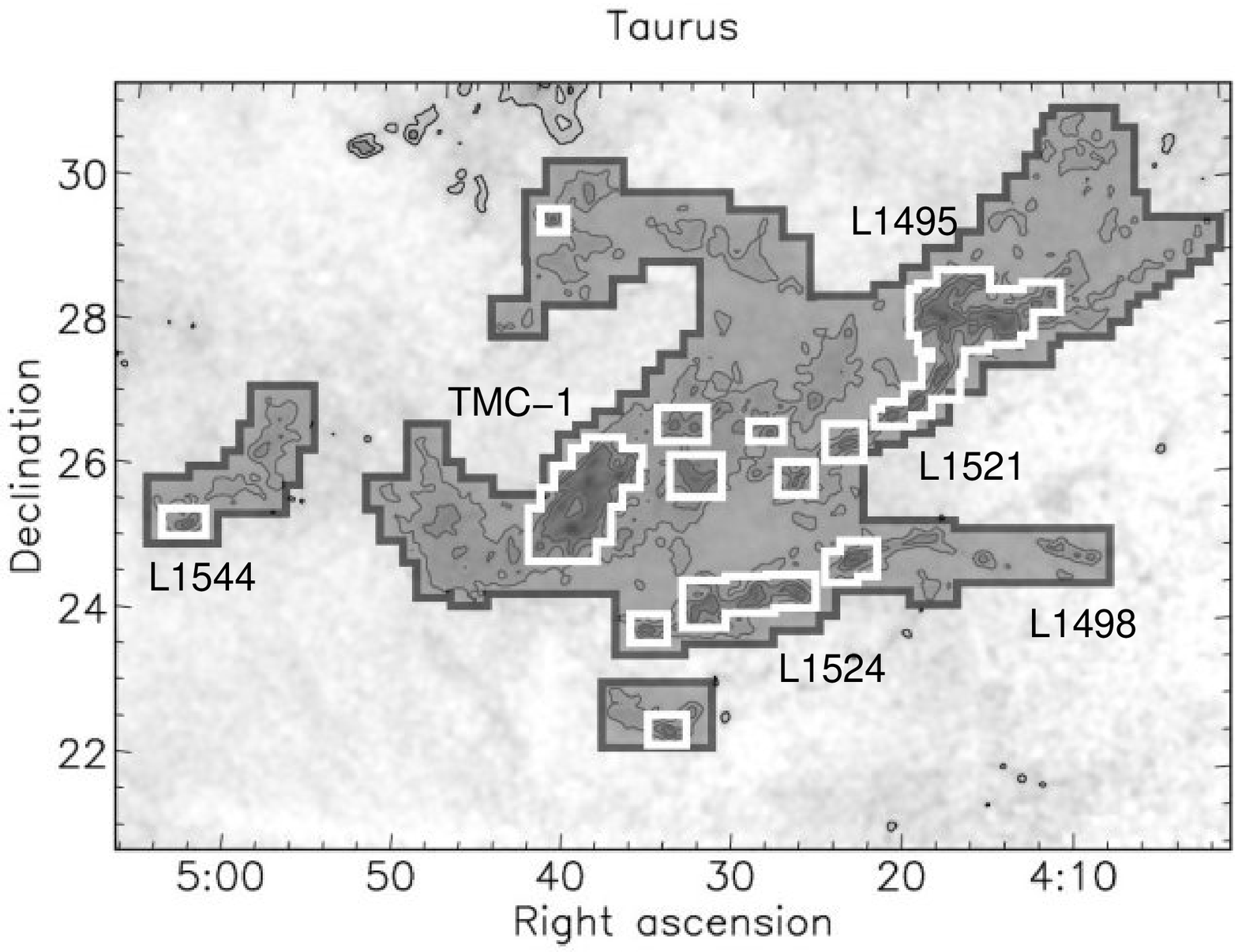}
\vspace*{20cm}
\end{picture}
\caption{}
\label{fig4}
\end{figure}

\newpage

\begin{figure}
\setlength{\unitlength}{1mm}
\begin{picture}(120,120)
\includegraphics{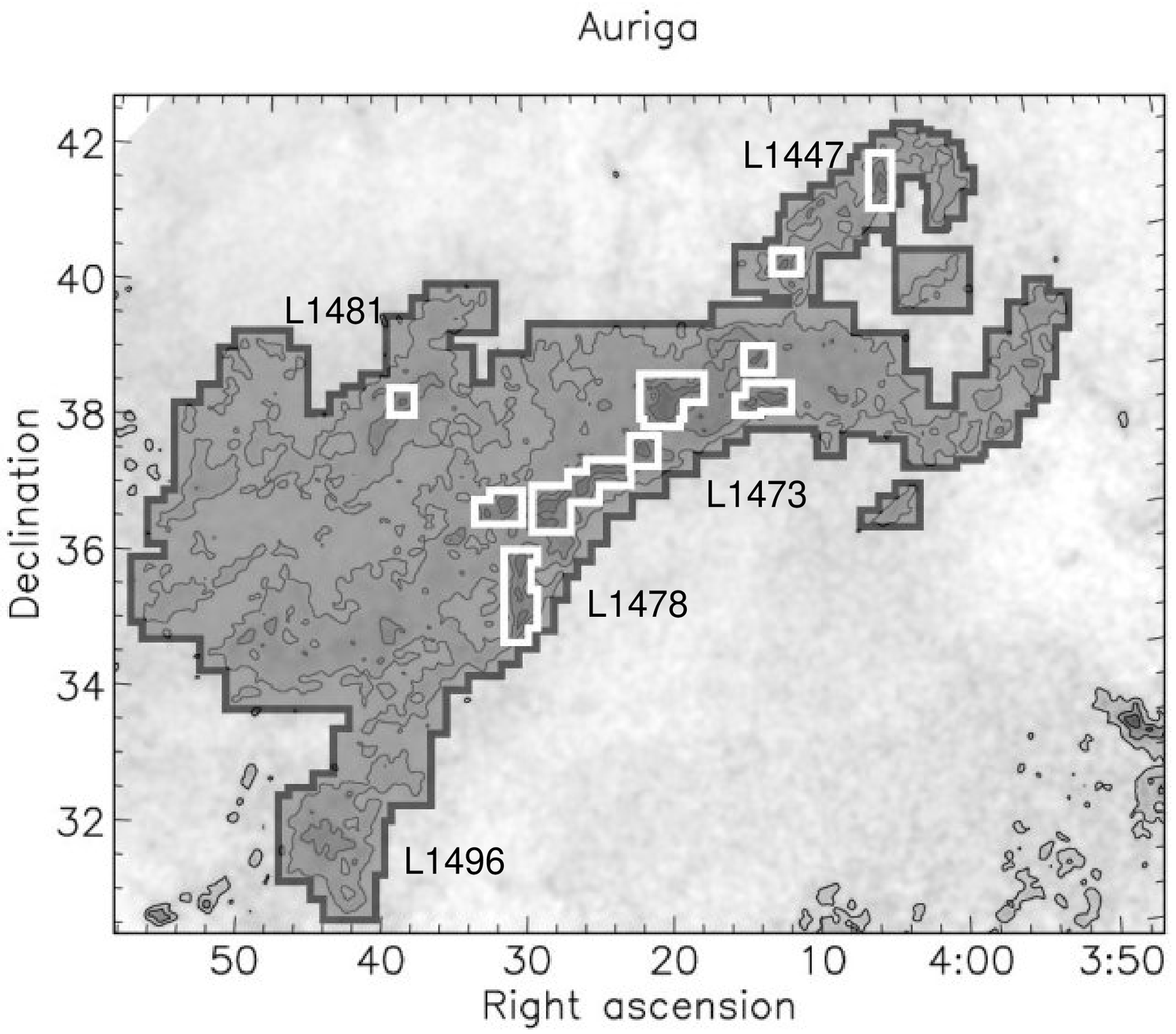}
\vspace*{20cm}
\end{picture}
\caption{}
\label{fig5}
\end{figure}

\newpage

\begin{figure}
\setlength{\unitlength}{1mm}
\begin{picture}(120,120)
\includegraphics{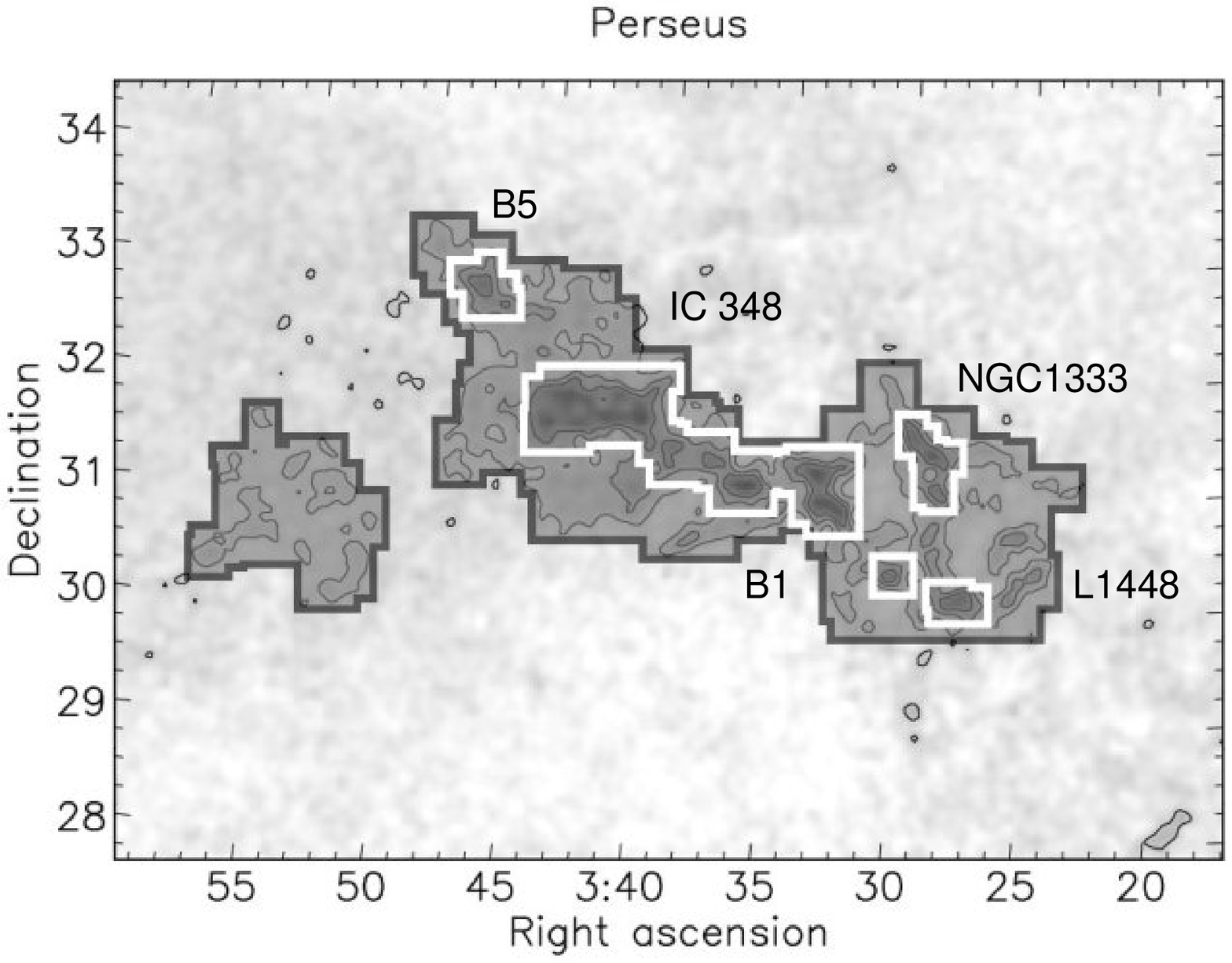}
\vspace*{20cm}
\end{picture}
\caption{}
\label{fig6}
\end{figure}

\newpage

\begin{figure}
\setlength{\unitlength}{1mm}
\begin{picture}(120,120)
\includegraphics{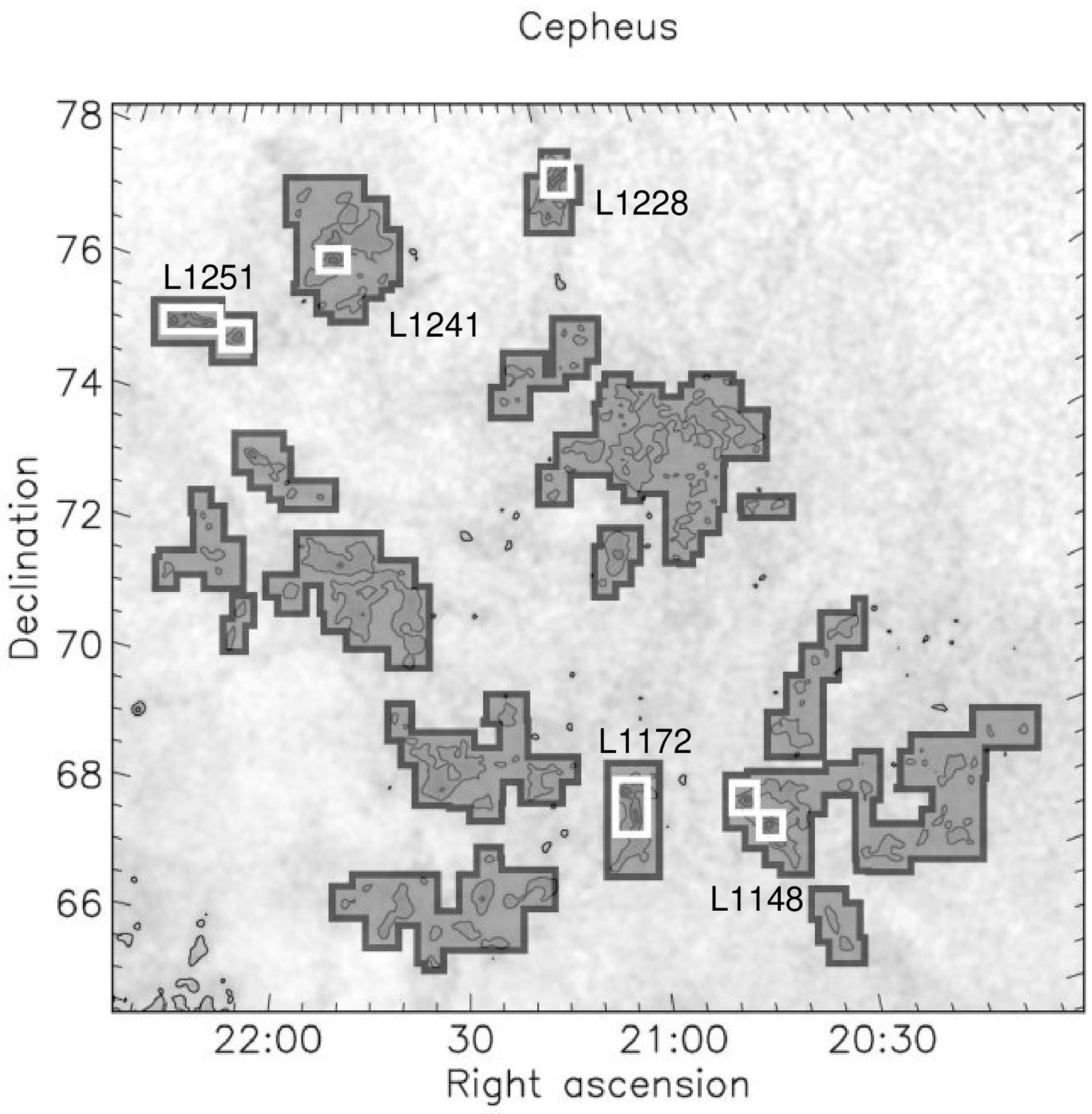}
\vspace*{20cm}
\end{picture}
\caption{}
\label{fig7}
\end{figure}

\newpage

\begin{figure}
\setlength{\unitlength}{1mm}
\begin{picture}(120,120)
\includegraphics{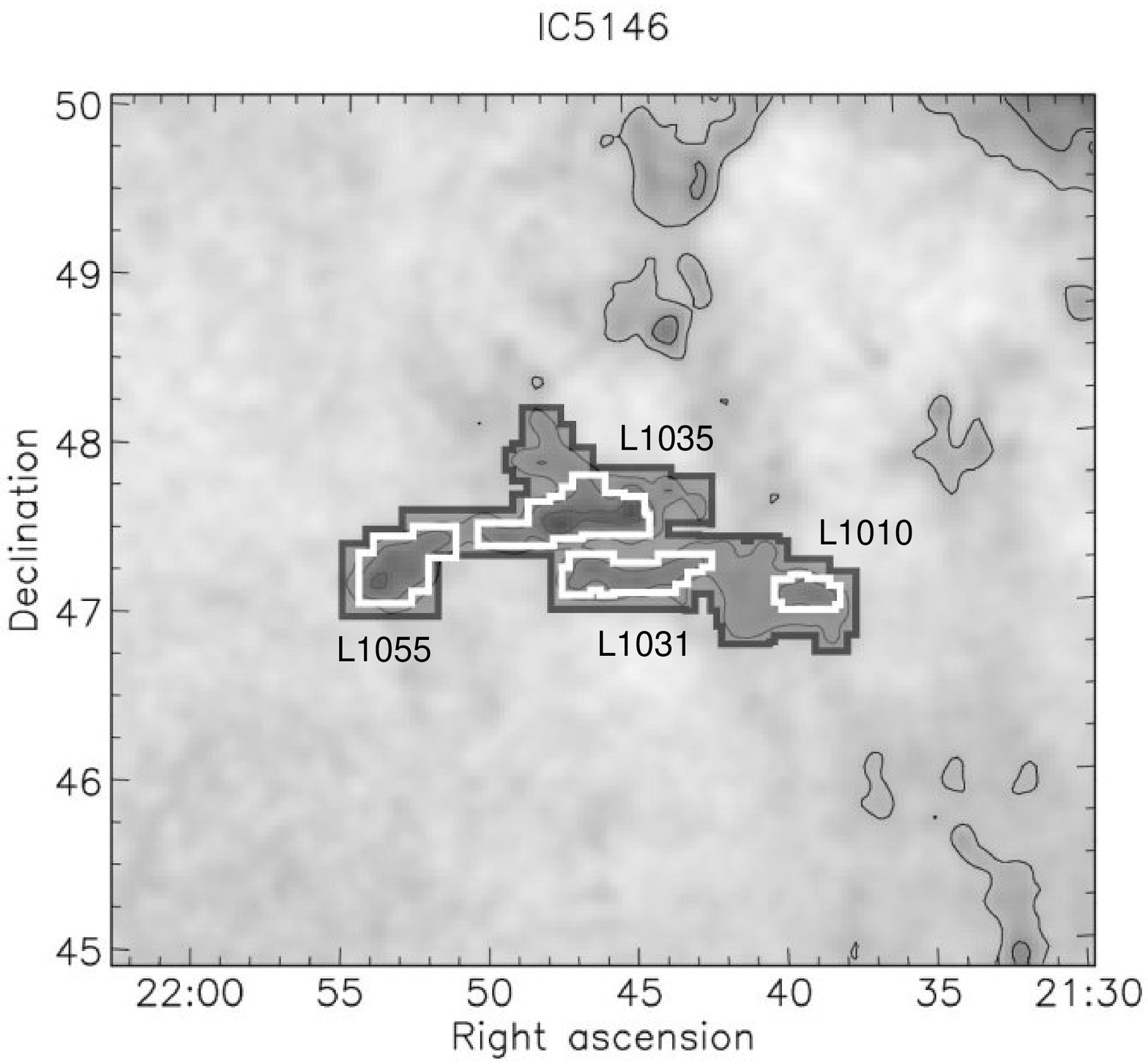}
\vspace*{20cm}
\end{picture}
\caption{}
\label{fig8}
\end{figure}

\newpage

\begin{figure}
\setlength{\unitlength}{1mm}
\begin{picture}(120,120)
\includegraphics{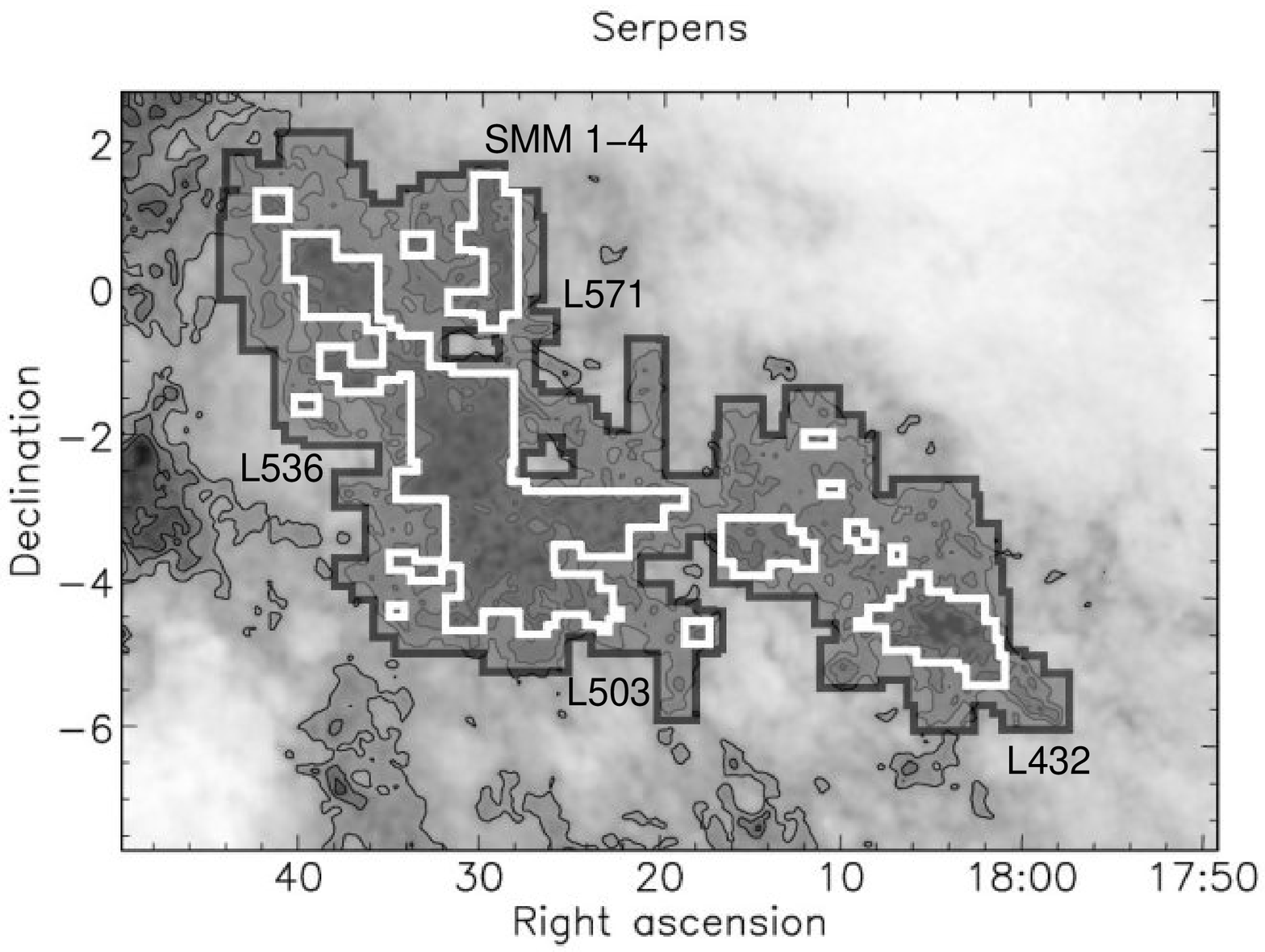}
\vspace*{20cm}
\end{picture}
\caption{}
\label{fig9}
\end{figure}

\newpage

\begin{figure}
\setlength{\unitlength}{1mm}
\begin{picture}(120,120)
\includegraphics{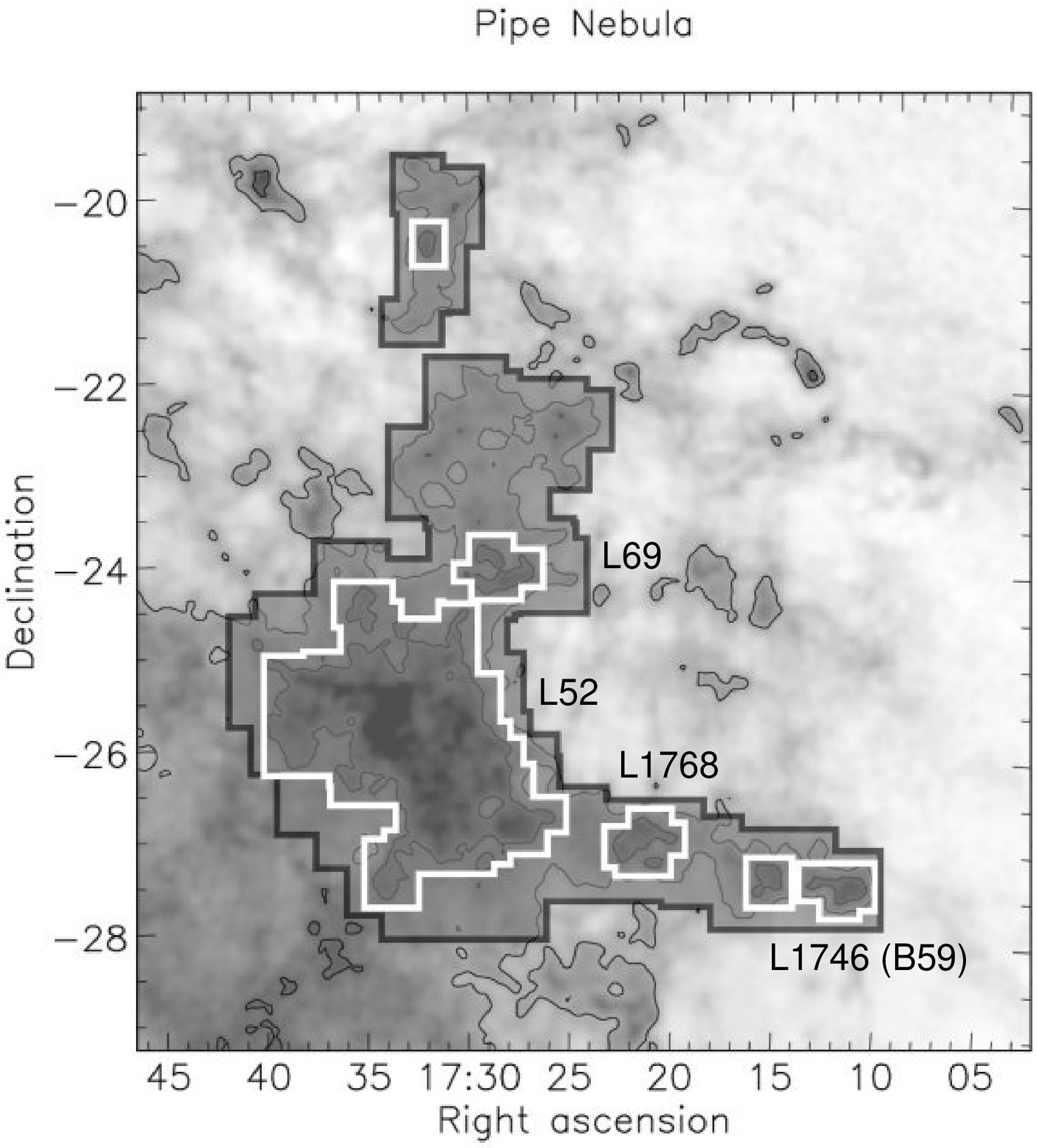}
\vspace*{20cm}
\end{picture}
\caption{}
\label{fig10}
\end{figure}

\newpage

\begin{figure}
\setlength{\unitlength}{1mm}
\begin{picture}(120,120)
\includegraphics{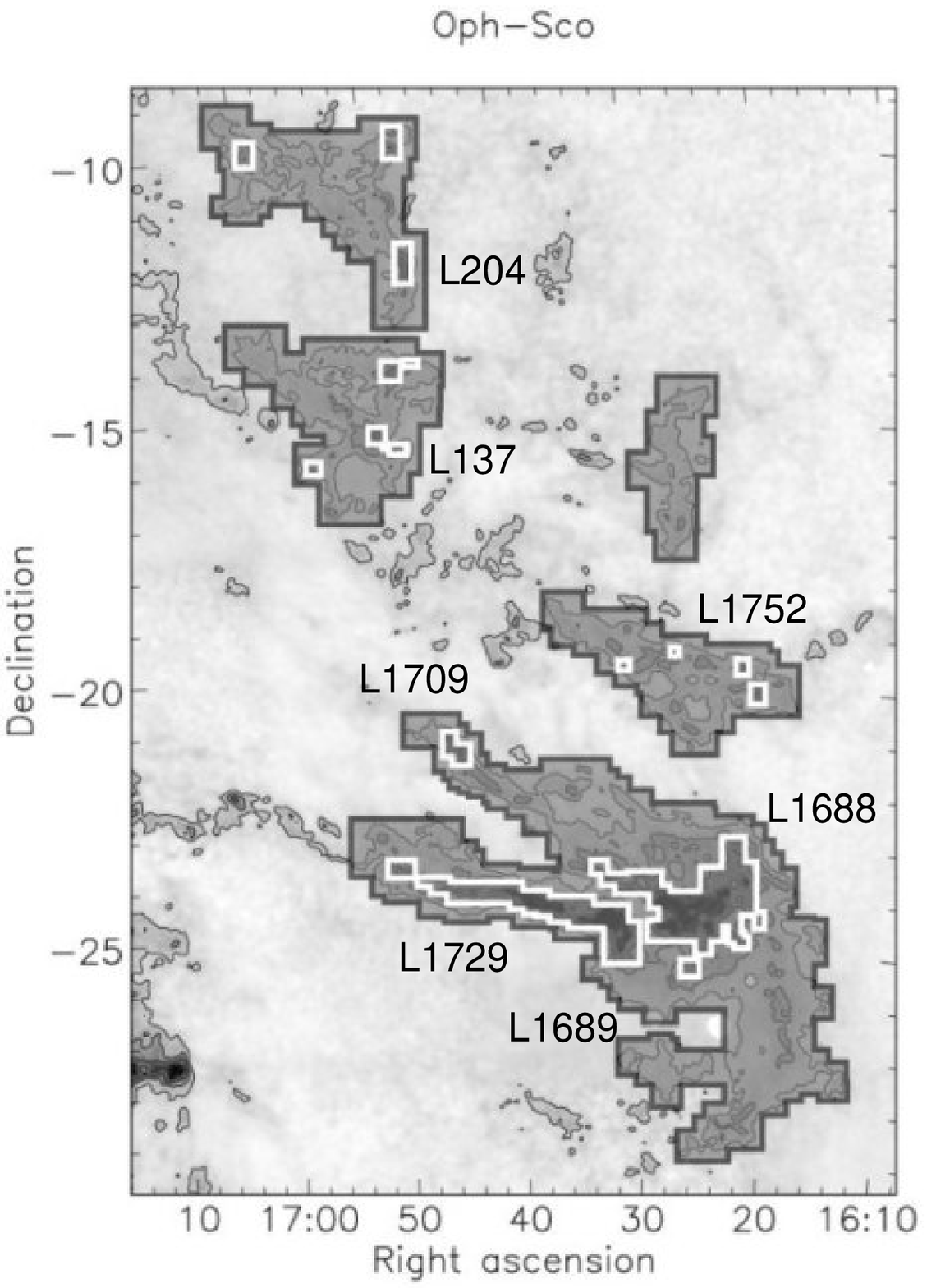}
\vspace*{20cm}
\end{picture}
\caption{}
\label{fig11}
\end{figure}

\newpage

\begin{figure}
\setlength{\unitlength}{1mm}
\begin{picture}(120,120)
\includegraphics{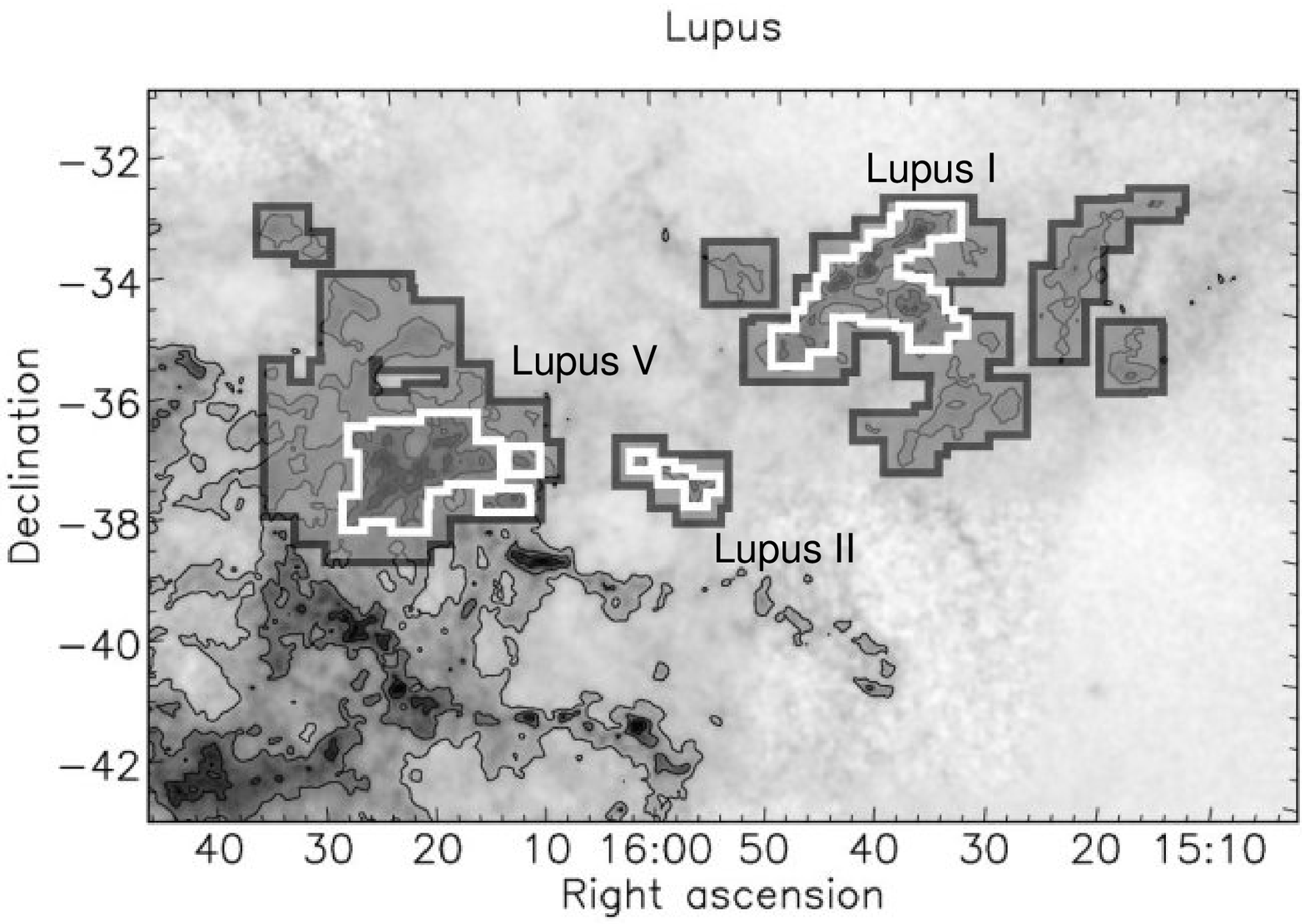}
\vspace*{20cm}
\end{picture}
\caption{}
\label{fig12}
\end{figure}

\newpage

\begin{figure}
\setlength{\unitlength}{1mm}
\begin{picture}(120,120)
\includegraphics{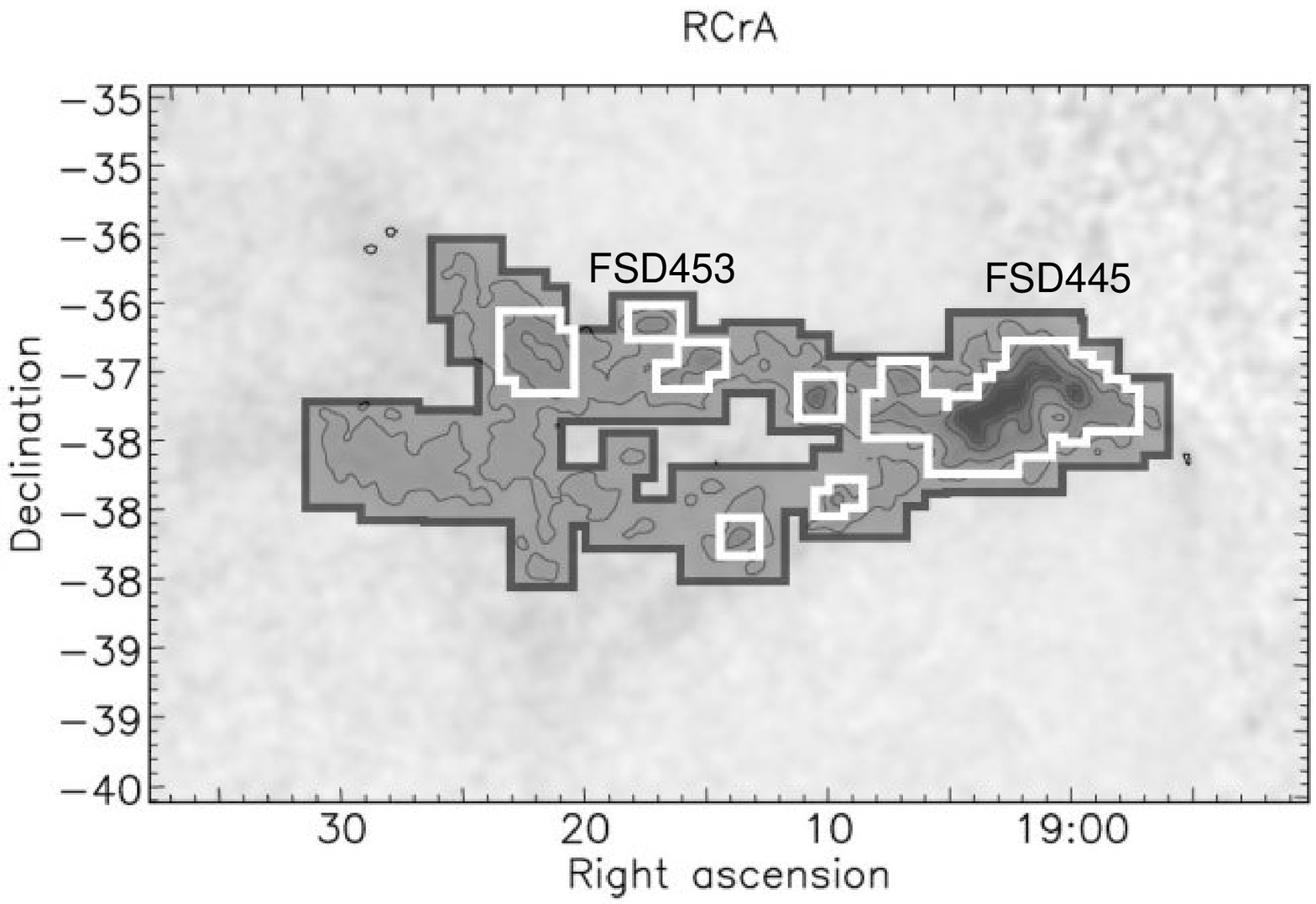}
\vspace*{20cm}
\end{picture}
\caption{}
\label{fig13}
\end{figure}

\newpage

\begin{figure}
\setlength{\unitlength}{1mm}
\begin{picture}(120,120)
\includegraphics{f14.eps}
\vspace*{20cm}
\end{picture}
\caption{}
\label{fig14}
\end{figure}

\newpage

\begin{figure}
\setlength{\unitlength}{1mm}
\begin{picture}(120,120)
\includegraphics{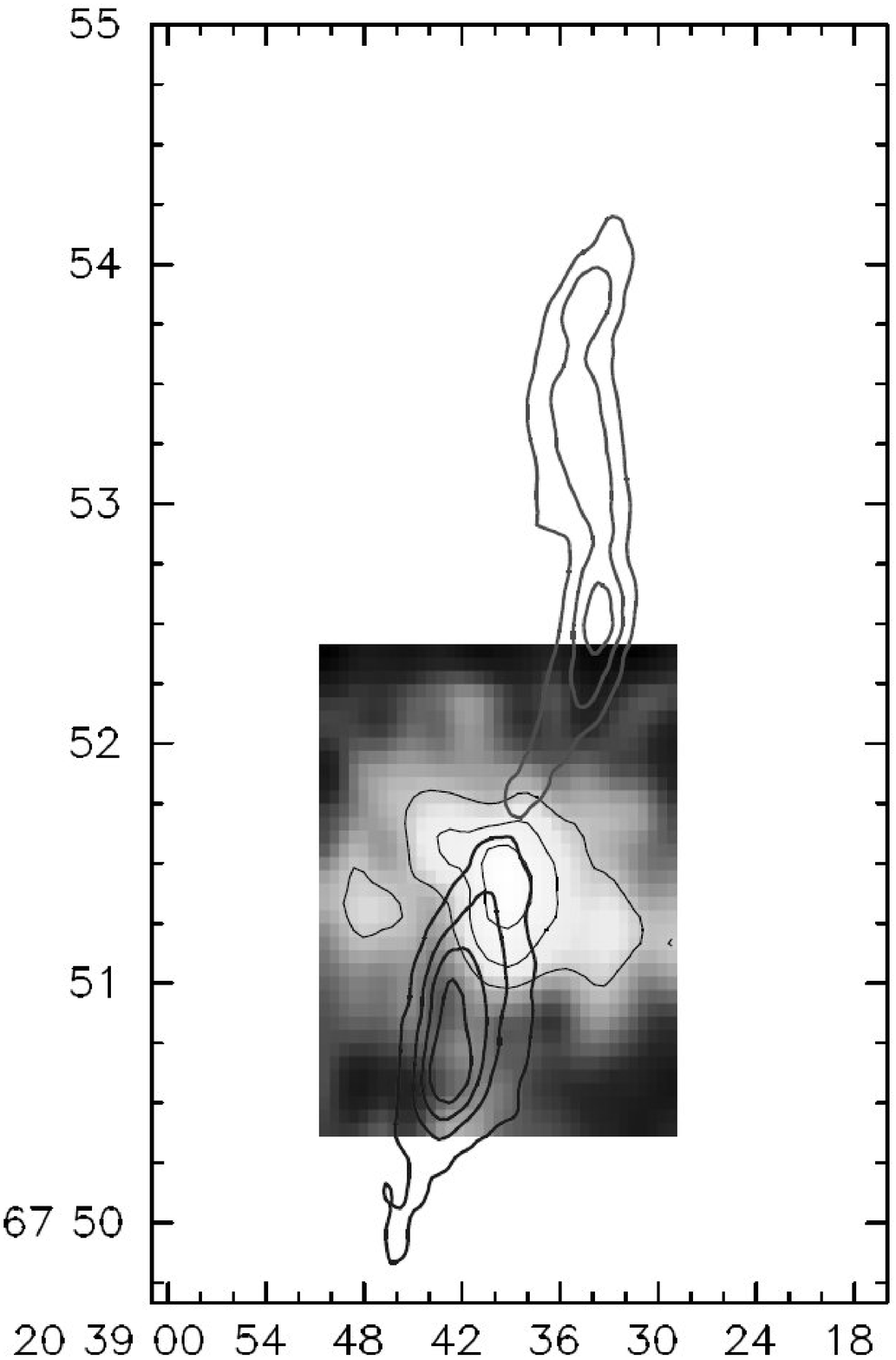}
\vspace*{20cm}
\end{picture}
\caption{}
\label{fig15}
\end{figure}

\newpage

\begin{figure}
\setlength{\unitlength}{1mm}
\begin{picture}(120,120)
\includegraphics{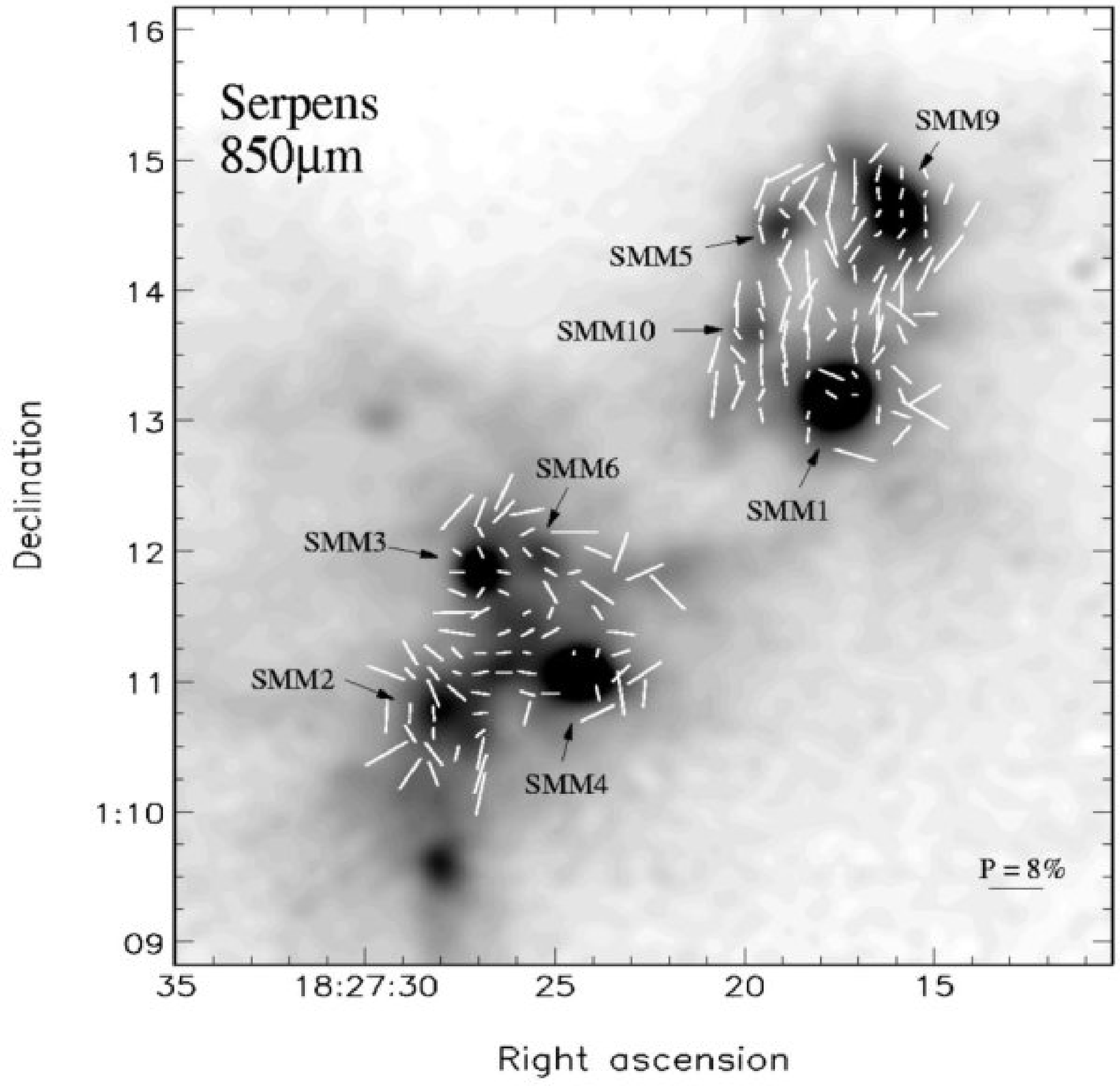}
\includegraphics{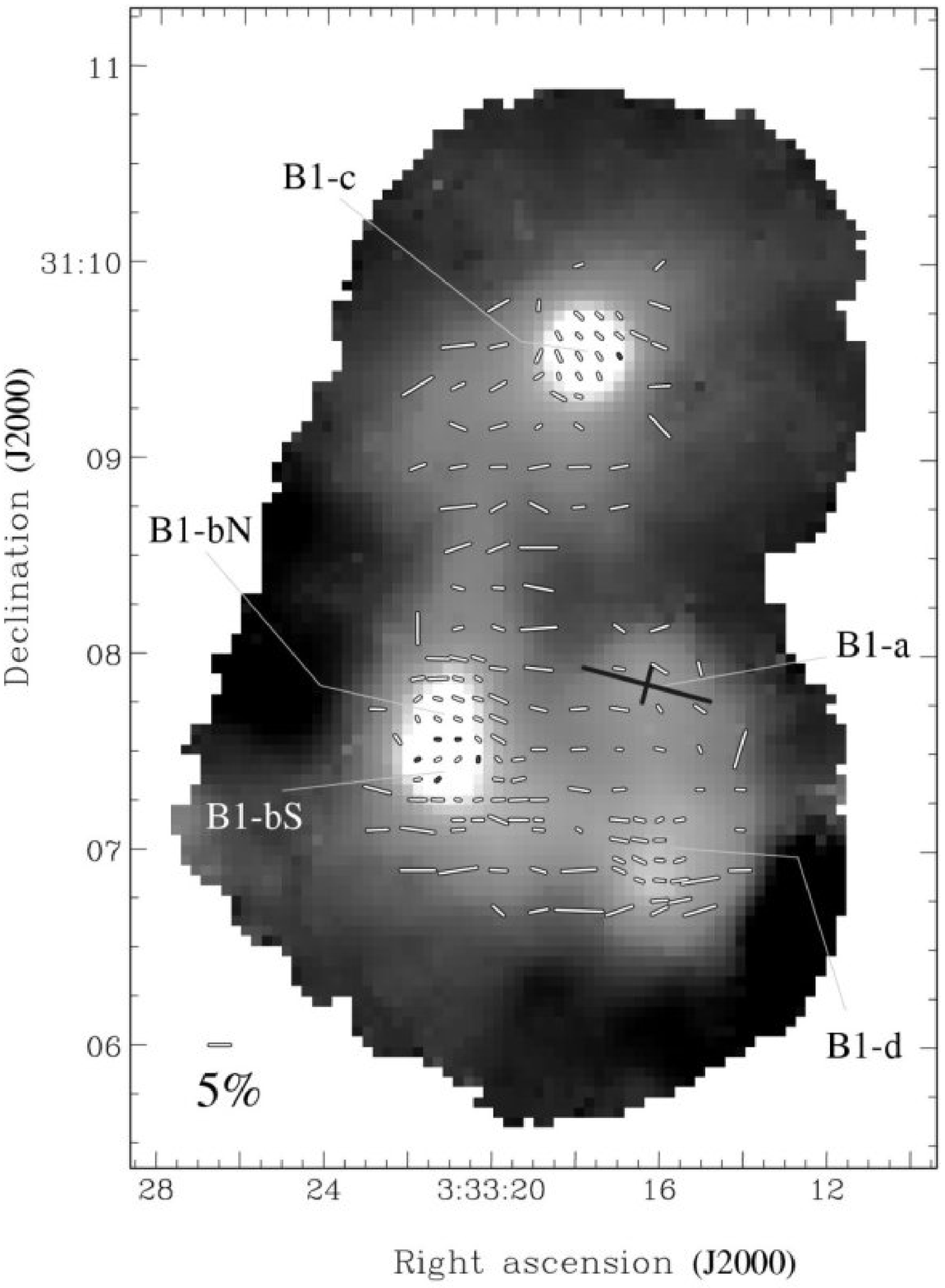}
\includegraphics{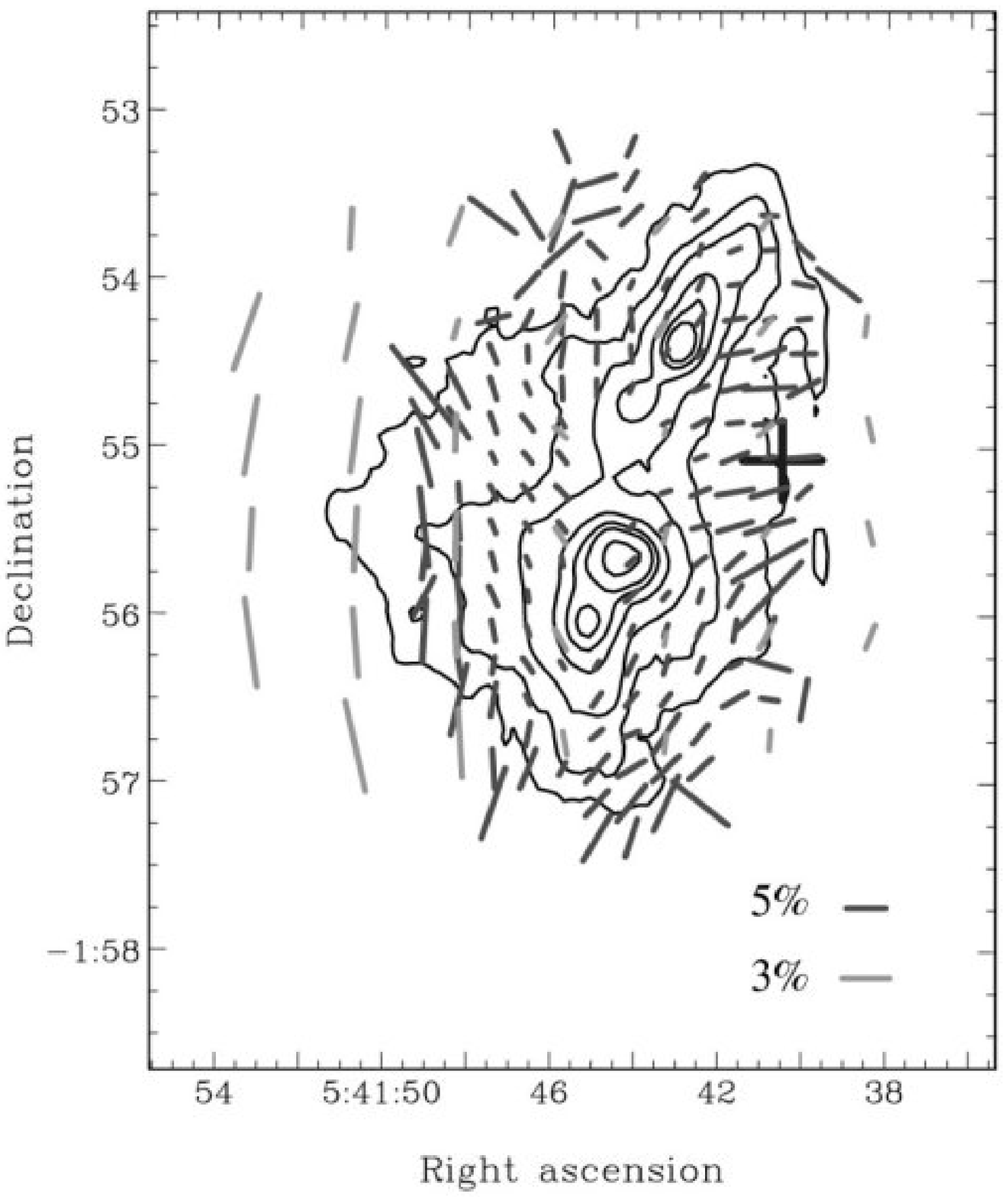}
\vspace*{20cm}
\end{picture}
\caption{}
\label{fig16}
\end{figure}

\end{document}